\documentclass[aps,pre,twocolumn,preprintnumbers,superscriptadress,
floatfix,nofootinbib]{revtex4-1}
\usepackage{natbib,hyperref}
\hypersetup{colorlinks=true, citecolor=blue, urlcolor=blue, linkcolor=blue}
\usepackage[english]{babel}
\usepackage[utf8]{inputenc}
\usepackage[T1]{fontenc}
\usepackage{graphicx}
\usepackage[caption=false]{subfig}
\usepackage{amsmath}
\usepackage{amsfonts}
\usepackage{amssymb}
\usepackage{color,soul}
\setulcolor{red}
\usepackage{xcolor}
\usepackage{multirow}
\begin{document}

\title{Violation of Stokes-Einstein and Stokes-Einstein-Debye
relations in polymers at the gas-supercooled liquid coexistence}

\author{Jalim Singh}
\altaffiliation{Present address: School of Physical Sciences, National
Institute of Science Education and Research, HBNI, Jatni, Bhubaneswar 752050,
India.}
\email{jalimsingh994@gmail.com}
\author{Prasanth P. Jose}
\altaffiliation{Corresponding author}
\email{prasanth@iitmandi.ac.in}
\affiliation{School of Basic Sciences, 
	Indian Institute of Technology Mandi,
Kamand, Himachal Pradesh 175005, India}


\begin{abstract} 

	Molecular dynamics simulations are performed on a system of model linear polymers to look at
	the violations of Stokes-Einstein (SE) and Stokes-Einstein-Debye (SED) relations near the mode
	coupling theory transition temperature $T_c$ at three (one higher and two lower) densities. 
	At low temperatures, both lower density systems show stable gas-supercooled-liquid coexistence
	whereas the higher density system is homogeneous. We show that monomer density relaxation 
	exhibits SE violation for all three densities, whereas molecular density relaxation shows a
	weak violation of the SE relation near $T_c$ in both lower density systems. This study
	identifies disparity in monomer mobility and observation of jumplike motion in the typical
	monomer trajectories resulting in the SE violations. In addition to the SE violation, a
	weak SED violation is observed in the gas-supercooled-liquid coexisting domains of the
	lower densities. Both lower density systems also show a decoupling of translational and
	rotational dynamics in this polymer system.  

\end{abstract}

\maketitle

\section{Introduction}

Simple liquids show coupling of translational diffusion coefficient $D$ 
and viscosity $\eta$ through the SE 
relation \cite{ham,j:stokes_elstc,j:einstein_brow} 
\begin{equation}
	\label{e:se}
	D = \frac{k_B T}{c\pi\eta r}, 
\end{equation}
where $k_B$ is the Boltzmann constant, $r$ is effective hydrodynamic radius 
of a Brownian particle immersed in the fluid. The constants $c=$ 4 and 6 in 
Eq. (\ref{e:se}) are respectively for slip and stick boundary conditions.
Thus, Eq. (\ref{e:se}) can be written as $D\eta/T=$ constant that is obeyed in the
liquids at high temperatures. A calculation of $\eta$ of a system is
computationally expensive in the supercooled liquids due to the large fluctuations
in their stress autocorrelation functions \cite{j:kawasaki_tmscl}. Therefore,
instead of viscosity $\eta$, the relaxation time $\tau$ is computed from the
simulations of glass-forming liquids (GFLs) with the approximation 
$\tau \propto \eta/T$ \cite{j:shiladitya,j:bhowmik_sk,s:deb_stil,j:onuki,senotes}. 
Deviations from $D\tau=$ constant are considered as the SE violations due to the 
decoupling in $D$ and $\tau$. These violations of the SE relation have been observed 
in simulations \cite{j:shiladitya,j:starr_fst,j:leporini,j:pan_SE} and 
experiments \cite{j:edigermd,j:leporini5,j:kazem,j:ganapathy_fst} of several supercooled 
GFLs. Earlier studies show that a possible reason for the violation of the SE
relation is dynamical heterogeneity (DH) in the supercooled liquids 
\cite{j:edigermd,b:berthierdh,j:szamel2}. There is a consensus among the
glass community that glassy systems lose homogeneity in the dynamics due to the
presence of mobile (fast-moving) and immobile (caged) particles \cite{j:scg4}, 
noticeably in fragile GFLs \cite{j:edigermd}. The mobile particles move faster than the 
average motion, while immobile particles move slowly in the system, leading to the 
DH \cite{j:edigermd,j:scg6,j:shd_richert,j:scg4,j:scg3,b:berthierdh,j:szamel2}.

Near the glass transition, $D$ is dominated by the mobile particles, whereas $\tau$ is 
governed by the immobile particles (as the majority of the particles are caged). Increment
in $\tau$ is not accompanied by the decrement in $D$ due to different mobilities in the 
system, causing a decoupling in $D$ and $\tau$, which results in the breakdown of the SE 
relation. It has been argued that the hopping of particles from the cages formed 
by their neighbors has a role in the violation of the SE relation in the supercooled
GFLs \cite{j:SEparisi,s:chong_hop,j:sun_SE-SED_Janus}. A study on supercooled
hard-sphere liquid by Kumar \textit{et al.} \cite{j:szamelsedse} shows that the SE 
relation violates due to the presence of mobile particles, whereas immobile particles
obey it. However, few recent studies show that both mobile and immobile particles violate 
the SE relation \cite{j:starr_fst,j:pan_SE}.  

In molecular GFLs, system shows violation of SED relation connecting orientational 
relaxation and viscosity of the liquid. The relaxation time of rotational 
correlation function $\tau_l$, defined using  $l$th order Legendre polynomial,   
is related to the viscosity $\eta$ via SED relation \cite{b:debye_book,bap,j:tarjus}
\begin{equation}
	\label{e:dse}
	\frac{1}{\tau_l} = \frac{l(l+1)k_B T}{6\eta V_h},
\end{equation}
where $V_h=(4/3)\pi R_h^3$ is the hydrodynamic volume of a molecule; $R_h$ is
the effective hydrodynamic radius of the molecules. Hydrodynamic radius $R_h$ and 
radius of gyration $R_g$ are unequal, in general. However, they are related in the 
Kirkwood approximation \cite{dae,deg,kirkwood}. Few recent studies, on the 
relation between $R_g$ and $R_h$ of the longer polymer chains, show that 
$R_g/R_h=a$ \cite{rac,douglas,dunbeg,douglas17}. Here $a$ is a 
constant depending on the molecular weight, topology, etc. of the polymers. 
In addition to this, Costigliola \textit{et al.} \cite{dyre_SE} and 
Ohtori \textit{et al.} \cite{ohtori15,ohtori18} revisit the Stokes-Einstein 
relation showing that $R_h\propto\rho^{{-1}/3}$, which modifies the relation as 
\begin{equation}
	\frac{D\eta}{\rho^{1/3}k_BT} = \textnormal{const}.
\end{equation}
This relation omits the importance of hydrodynamic diameter and rely instead on the 
system density $\rho$; these authors examined the relation in Lennard-Jones 
liquids above the critical density. The coupling between viscosity $\eta$ 
and rotational correlation time $\tau_l$ of molecular liquids holds the SED relation at high
temperatures, whereas it violates in the supercooled liquids and 
glasses \cite{j:stanley_SE-SED-Water,j:leporini5,j:ngai_DSE}. Using this approximation, 
Eq. (\ref{e:dse}), and Eq. (\ref{e:se}) suggest that $D\tau_l=$ constant in molecular
liquids at high temperatures \cite{j:tarjus}. Thus, the translational molecular diffusion 
$D$ is coupled to the rotational relaxation time $\tau_l$ of the molecules. However, 
simulation studies by Michele \textit{et al.} \cite{j:micheleleporini1,j:micheleleporini2}
show translational and rotational jump dynamics leads to violation of SE and SED
relations in a system of diatomic rigid-dumbbell molecules.

Extending such studies (performed on binary mixtures and dumbbells) to polymers is a 
daunting task due to the complexity of the system. Identification of jumplike motions are
reported in the continuous-time random walk simulation of 
Helfferich \textit{et al.} \cite{j:helfrich1} with a special definition of the jumps in a
supercooled short-chain melt. Another study by 
Pousi \textit{et al.} \cite{j:leporini,j:leporini6} argued that picosecond dynamics of the 
caged particles corresponding to short time ($\beta-$relaxation), is related to the 
violation of SE relation in supercooled linear polymers. A very recent study shows that the
violation of SE relation is related to the structural changes in the supercooled linear
polymer melt \cite{j:sepoly}. Therefore, due to the molecular complexity, it is strenuous to
examine the SE and SED violation in polymer liquids and to identify a possible reason
compared to the simulation studies of atomic GFLs. This study attempts a direct identification
of  SE and SED violations in a simple polymer model that is known to undergo glass transition 
and relates it to the known reasons that are manifestations of DH in atomic GFLs.

Polymers are extended molecules that relax by collective rearrangements of monomers.
Therefore, identification of the mobile and immobile particles is 
expected to be easier in low density where there is more free-space available at lower
temperatures. Polymer systems with attractive interactions in the lower density show
coexistence of dilute gas and supercooled liquid domains at lower temperatures with cavities.
On the surface of these cavities, there is free space available for chains leading to a large
variation in mobilities, which can show direct evidence of SE and SED violations and their 
possible microscopic origins. In this study, we examine the violations of the SE and SED 
relations in a linear polymer system at number densities $\rho=$ 1.0 \cite{j:sastryfvol},
0.85 \cite{j:bpk}, and 0.7, which we call as one higher and two lower (relatively) density
systems, from $T=$ 2.0 to near their respective $T_c$ (in unequal grids of temperature),
i.e., 0.36 for the higher density and 0.4 for both lower densities. Both lower density systems
are quenched to temperatures beyond the spinodal limit of stability, which phase separate via 
spinodal decomposition \cite{huang} resulting in stable coexistence of (dilute) gas and 
supercooled-liquid with long equilibration times. Details of non-equilibrium dynamics of 
cavity formation are given in earlier 
studies \cite{j:foffi,j:frederic,j:godfrin,j:pinaki2,j:testard1}.
Recently, Priezjev and Makeev examined an effect of shear strain on the porous structure in 
a model binary glass in non-equilibrium \cite{makeev1,makeev2}.
In our study, monomer relaxation shows SE violation in all three systems; both lower density 
systems show pronounced violation that is attributed to the enhanced disparity in the motion of 
the monomers that arises due to the surface of the cavities. A pronounced violation of SE relation
is due to both the mobile and immobile particles, which agrees with the earlier studies on violation
of SE and SED relations in atomic model GFLs \cite{j:starr_fst,j:pan_SE}. The SED relation is 
obeyed in the higher density system, whereas it is weakly violated in both lower density systems. 

The rest of the manuscript is organized as follows: Sec. \ref{ds} gives the details of
the simulations. Results are presented in subsections of \ref{rs}. Subsec. \ref{stb} 
present stability analysis of the phase coexistence of gas-supercooled-liquid. The study of 
polymer relaxation and diffusion is presented in Subsec. \ref{se:polyrel}, SE violations 
are discussed in Subsec. \ref{se:sev}, the distribution of particles' mobility and jumplike 
motions are presented in Subsec. \ref{se:mob}, and violation of SED relation and orientational 
relaxation are detailed in Subsec. \ref{se:ori}. Finally, Sec. \ref{se:sac} presents conclusions 
and a short summary. 

\section{\label{ds}Simulation details}

We simulate a system of $N_c=$ 1000 fully-flexible linear polymer 
chains, consisting of $n=$ 10 beads in each. Thus, the system consisting of $N=$ 10000
number of monomers is simulated at constant number density of monomers, i.e., 
$\rho =$ 0.7, 0.85, and 1.0. Inter-particle interactions are
modeled by truncated and shifted Lennard-Jones (LJ) potential, defined in terms of the 
particle diameter, $\sigma$ and the depth of the potential well, $\epsilon$ as
\begin{equation}\label{LJ}
	V_{LJ}(r)=4\epsilon\left[\left(\frac{\sigma}{r}\right)^{12}
	-\left(\frac{\sigma}{r}\right)^6\right] - V_{LJ}(r_c).
\end{equation} 
Here, the LJ potential cut-off is $r_c = 2 \times 2^{1/6}\sigma$. The bond connectivity between
consecutive monomers along a chain is modeled by the LJ and
finitely extensible non-linear elastic (FENE) potentials \cite{j:kremer,j:bpk}; the FENE 
potential is given in terms of $R_0$, the maximum displacement between a pair of consecutive 
monomers, and the elastic constant $k_0$ as

\begin{equation}
	V_{FENE}(r)=\begin{cases} -\frac{1}{2}k_0R^2_0 \ln\left(1-\frac{r^2}{R^2_0}\right)
		-E_b  & 0 < r< R_0\\
		\infty & r \geq R_0,
	\end{cases}
\end{equation}

where $k_0=30\epsilon/\sigma^2$ and $R_0=1.5\sigma$. The equations of motion are integrated 
using the velocity Verlet algorithm \cite{aat} with time step $\Delta t=$ 0.002 
for $\rho=$ 0.85, and $\Delta t=$ 0.0001 for $\rho=$ 1.0 and 0.7 system. We prepare the 
linear polymer melt at temperature $T=$ 8.0, for the respective density, in the
microcanonical (constant NVE) ensemble. The equilibrated configurations at $T=$ 8.0 are used 
as the initial configurations for the temperatures $T=$ 2.0--0.36. 
Long simulation time before the production runs ensures fluctuation of the system around the 
average target temperature during the data collection in constant NVE ensemble and follow a
criterion of the time required for the relaxation of end-to-end vector below $10\%$ 
of its initial value at all temperatures \cite{j:helfrich1}. All the quantities presented
in this work are in LJ units, i.e., the length is expressed in terms of the bead 
diameter $\sigma$, the number density as $\sigma^{-3}$, the temperature as $\epsilon/k_B$, 
and the time as $ \sqrt{m \sigma^2 /\epsilon}$. 

\section{Results and discussions \label{rs}}

\subsection{Steady-state density relaxation in gas-supercooled-liquid domains \label{stb}}
\begin{figure}
	\includegraphics[width=8.4cm, height=4.1cm]{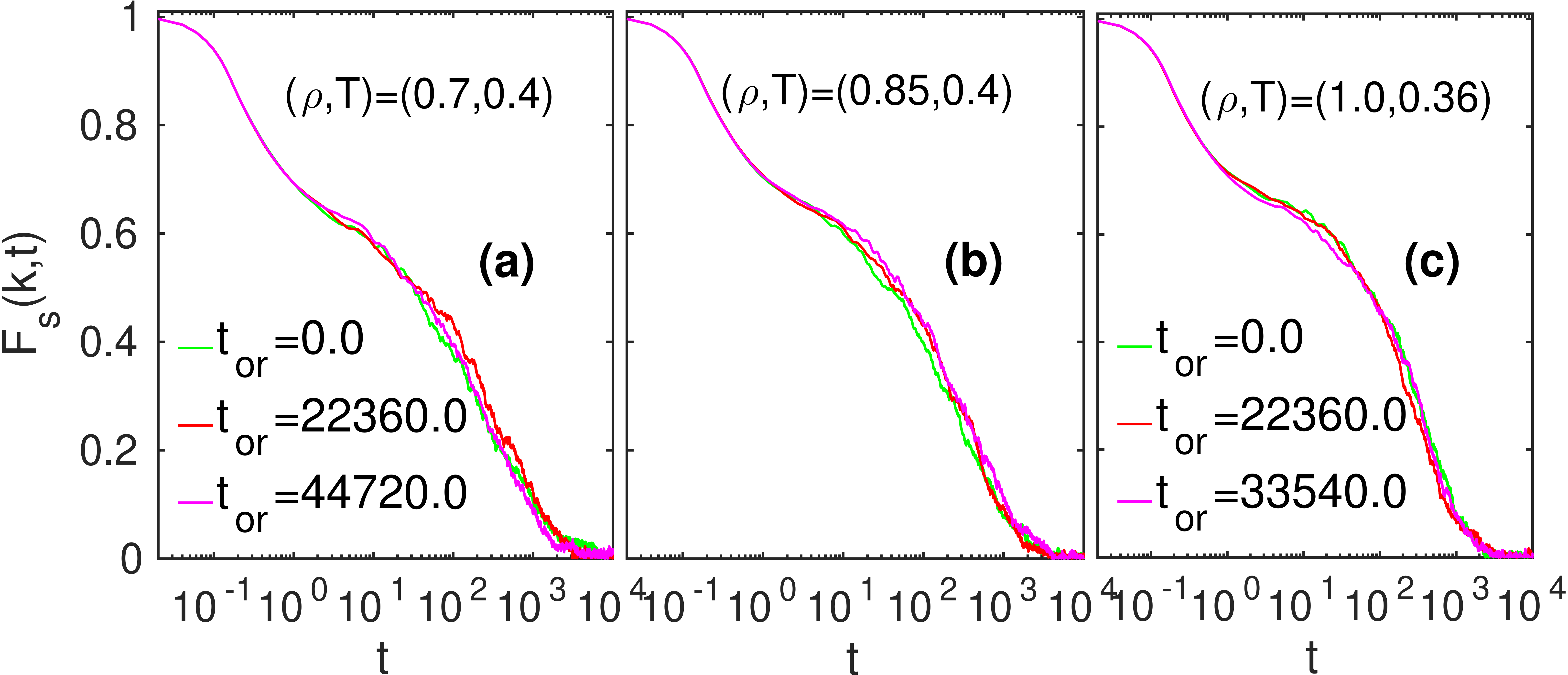}
	\caption{\label{f:fsktss} Incoherent intermediate scattering function, $F_s(k,t)$ of 
	monomers is calculated at different time origins of the low temperatures.
	The legend in (b) is same as in (c).}
\end{figure}

State points corresponding to decreasing order of temperatures in both lower density systems
show coexistence of gas and liquid at intermediate temperatures, e.g., $T=0.7$ for 
$\rho=$ 0.85, and $T=$ 1.0 for $\rho=$ 0.7 (see Fig. \ref{f:msdmap} and Fig. \ref{f:pnb}).
However, the system shows the phase-coexistence of dilute gas and supercooled liquid at
lower temperatures. In the moderately supercooled state, the system attains its steady-state 
where the cavities are stable, whereas the cavities at the liquid-gas phase coexistence move, 
thus the system attains density relaxation. Before presenting the glass transition studies, we
look at the single-particle and collective density relaxation to ensure the ergodicity of the 
system. Studies on aging compare the time-origin dependent density-relaxation using incoherent
intermediate scattering function $F_s(k,t)$ \cite{kob_aging}, which show different relaxation
time for different time origins. Therefore, we look into the relaxation of density fluctuations
in $F_s(k,t)$ \cite{ham} at time origins longer than the $\alpha-$relaxation
time (presented later in this paper), which is defined as
\begin{equation}
	\label{e:fskt}
	F_s(k,t)= \frac{1}{N} \left \langle \rho^s_{\mathbf{k}}(t)\rho^s_{-\mathbf{k}}(0)
	\right \rangle,
\end{equation}
where $\rho^s_{\mathbf{k}}(t)= \exp[-i \mathbf{k} \cdot \mathbf{r}_i(t)]$. Here, 
$k$ is the wavenumber and $\mathbf{r}_i$ is the position vector of an $i^{th}$ particle.
The $F_s(k,t)$ shown in Fig. \ref{f:fsktss} is calculated at a wavenumber $k$ that 
corresponds to the first peak of static structure factor $S(k)$; $k=$ 6.9
for $\rho=$ 0.85 and $k=$ 7.1 for $\rho=$ 0.7 and $\rho=$ 1.0. The $\alpha-$relaxation
time of the monomers $\tau_\alpha = \int_0^{\infty} F_s(k,t) dt $
gives the relaxation time of transient cages formed by their neighbouring particles.
The time difference between two time origins is chosen to be greater than $\tau_\alpha$
(see Fig. \ref{f:talffit} for $\tau_\alpha$) to look at the average density relaxation of 
monomers in that time window. The $F_s(k,t)$ of all three densities at the lowest 
temperatures, given in Figs. \ref{f:fsktss}(a--c), does not show any systematic variation
in the nature of density relaxation with time, at different time origins of the
correlation, therefore, they are independent of the time origin. A small difference due
to fluctuations in the tail of $F_s(k,t)$ in lower densities is similar to that at the
higher density.

\begin{figure}
	\includegraphics[width=7.6cm, height=6.8cm]{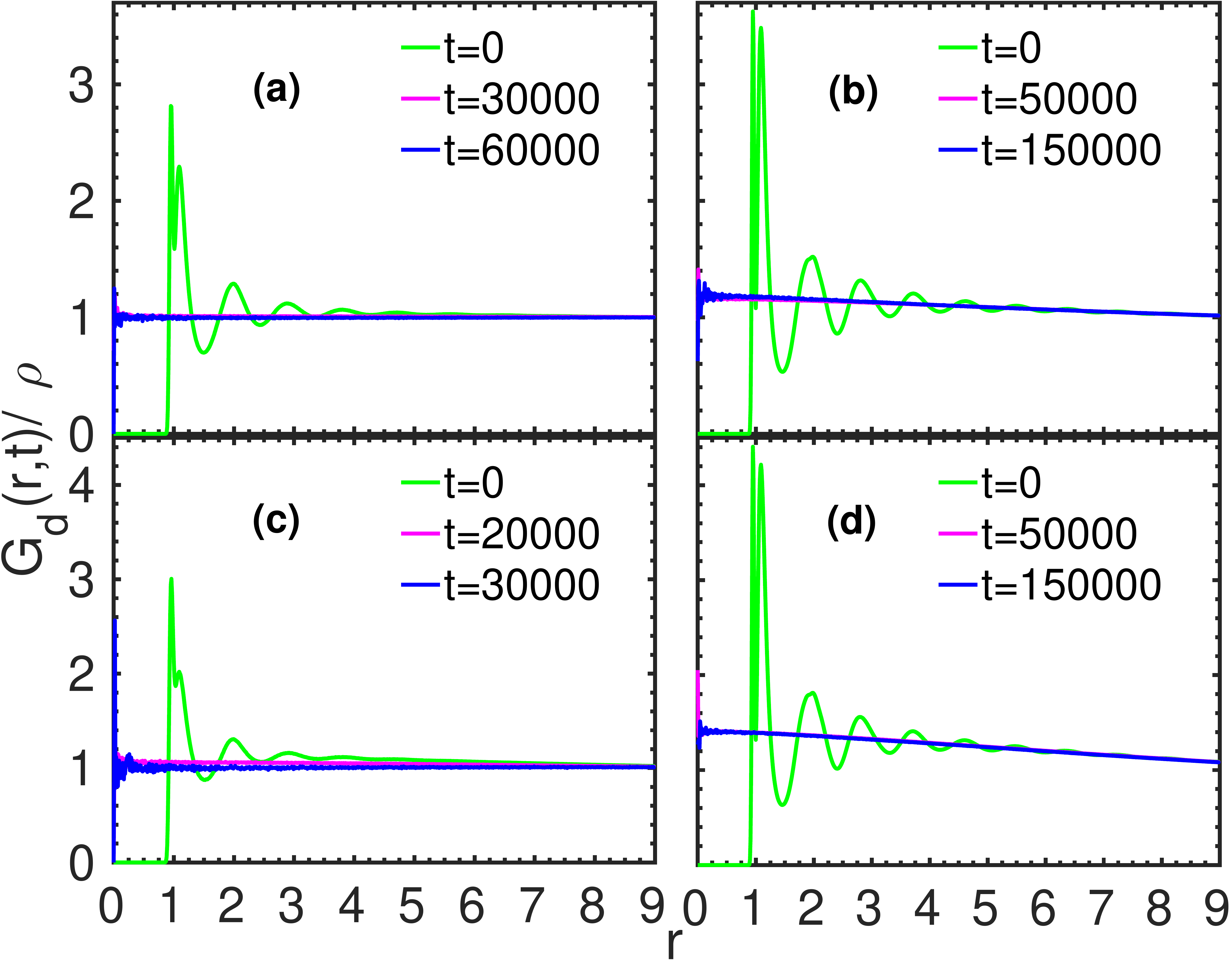}
	\caption{\label{f:gdrt} Distinct part of van-Hove correlation function scaled with the 
	density is plotted in (a) $\rho=$ 0.85, $T=$ 0.7; (b) $\rho=$ 0.85, $T=$ 0.4;
	(c) $\rho=$ 0.7, $T=$ 1.0; (d) $\rho=$ 0.7, $T=$ 0.4.}
\end{figure}

Collective relaxation dynamics of the phase-coexisting system is examined by 
distinct part of van-Hove correlation function, $G_d(r,t)$ \cite{ham},  
\begin{equation}
	\label{e:gdrt}
	G_{d}(\mathbf{r},t) = \frac{1}{N} \left\langle {\sum\limits_{i=1}^{N}\sum\limits_{j=i+1}^{N}
	{\delta[\mathbf{{r}}+{\mathbf{r}}_{i}(0)-{\mathbf{r}}_{j}(t)]}} \right\rangle,
\end{equation}
that shows collective structural relaxation in different coordination shells with respect to
a reference particle at $t=0$. Figure \ref{f:gdrt} shows the $G_d(r,t)/\rho$ of the monomers at
one low and one high temperature of both lower densities at different times starting 
from $t=0$. At $t=$ 0, $G_d(r,0)/\rho=g(r)$ is showing oscillations with mean above 1.0, which
is due to the phase separation in the system. The neighboring particles around a tagged particle
move from their initial positions (as time progresses), leading to decay of  the $G_d(r,t)$
to 1.0 as in a homogeneous system ($T=$ 1.0). However,  $G_d(r,t)/\rho$  remains above 1.0 
for both lower density  systems that show coexistence of dilute gas and supercooled liquid
domains at low temperatures even at longest time scale in this study. Interestingly, 
Fig. \ref{f:gdrt} shows that the $G_d(r,t)$ of both lower densities coincides at
times $t=5\times10^3$ and $t=1.5\times10^5$, which shows that the shape and location
of the cavities are stable and do not change within our simulation time in the moderately
supercooled state. It is now interesting to look at polymer relaxation dynamics in the
higher and both lower density systems as temperature reduces.

\subsection{\label{se:polyrel} Polymer relaxation and diffusion }

\begin{figure}
	\includegraphics[width=7.3cm, height=5.8cm]{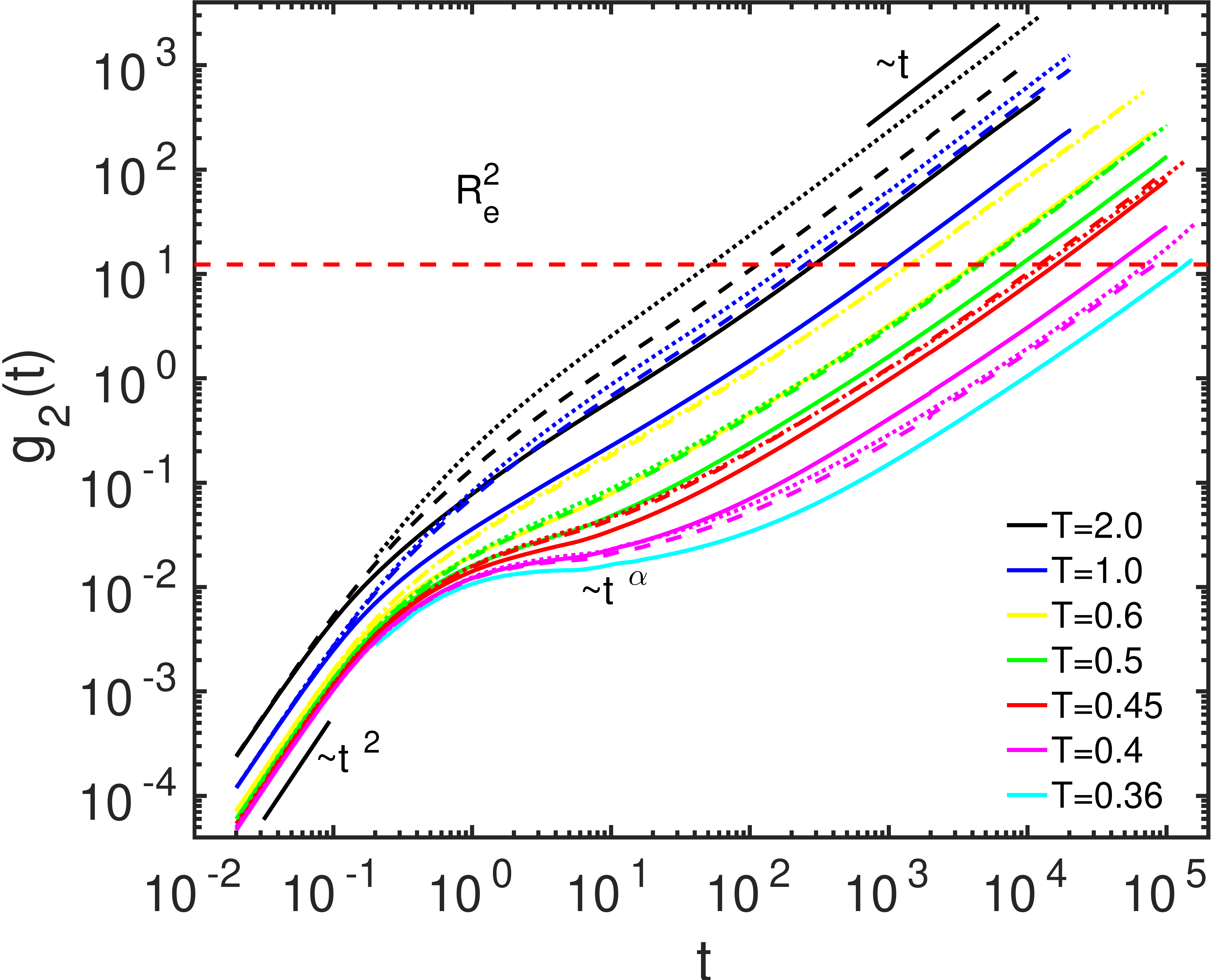}
	\caption{\label{f:cmsd} Center of mass MSD of chains, $g_2(t)$ is plotted against
	time $t$. Solid, dashed, and dotted lines are corresponding to the density 
	$\rho=$ 1.0, 0.85, and 0.7 respectively. At temperatures $T=$ 2.0--0.45, $g_2(t)$ of
	the higher density system is slower than both lower density systems. However, both 
	lower density systems show slow molecular movement than the higher density system
	from $T=$ 0.4 to below. At low temperatures ($T=$ 0.5--0.4), the sub-diffusive 
	regime ($g_2(t)\sim t^\alpha$) becomes pronounced due to the molecular cages, which
	is separated by the short time ballistic ($g_2(t)\sim t^2$) and long time 
	diffusive ($g_2(t)\sim t$) regimes. }
\end{figure}

The self-diffusion coefficient $D$ of the monomers is calculated from their
mean-squared displacement (MSD) as
$D=\lim\limits_{t \to \infty} {\langle [ {\mathbf{r}}(t)-\mathbf{r}(0)]^2 \rangle}/{6t}$.
Near the glass transition of unentangled polymer melts, the monomer MSD takes longer time 
to attain diffusive regime due to the chain connectivity \cite{j:barratrev}. Therefore, the 
monomer MSD shows a power law dependency as
${\langle [ {\mathbf{r}}(t)-\mathbf{r}(0)]^2 \rangle} \sim t^{0.63}$, 
before the starting of its diffusion \cite{j:barratrev}. However, the MSD of the
polymer molecules (center of mass) shows the diffusion at early times, though they are 
slow at short time and coincides with the monomer MSD at the long
times $t\simeq10^6$ \cite{j:barratrev,j:chong_fuchs}. Therefore, we calculate the self diffusion
coefficient from the center of mass MSD --- when the exponent $\alpha=$ 1.0 in the relation 
$g_2(t)=\langle[\mathbf r_{cm}(t)-\mathbf r_{cm}(0)]^2\rangle = 6Dt^\alpha$, 
where $t>10^5$ in the LJ units at the lowest temperatures (see Fig. \ref{f:cmsd}).
At this time scale, the molecular MSD also crosses the average squared end-to-end 
distance, i.e., $g_2(t) > R^2_e$, which is indicated by the red line at
$g_2(t)=R^2_e$ in Fig. \ref{f:cmsd}. Using the approximation
$\tau\propto\eta/T$ (as discussed above), the fractional SE relation reads
$D\sim\tau^{-\xi}$ ($0<\xi<1$), which shows that the diffusion coefficient and 
relaxation time (or viscosity) decouple from the usual SE relation, i.e., Eq. \ref{e:se},
which is also reported in various earlier studies of supercooled liquids and 
glasses \cite{j:szamelsedse,j:szamel2,j:starr_fst,j:ganapathy_fst}. 

\begin{figure}
	\includegraphics[height=3.8cm, width=8.8cm]{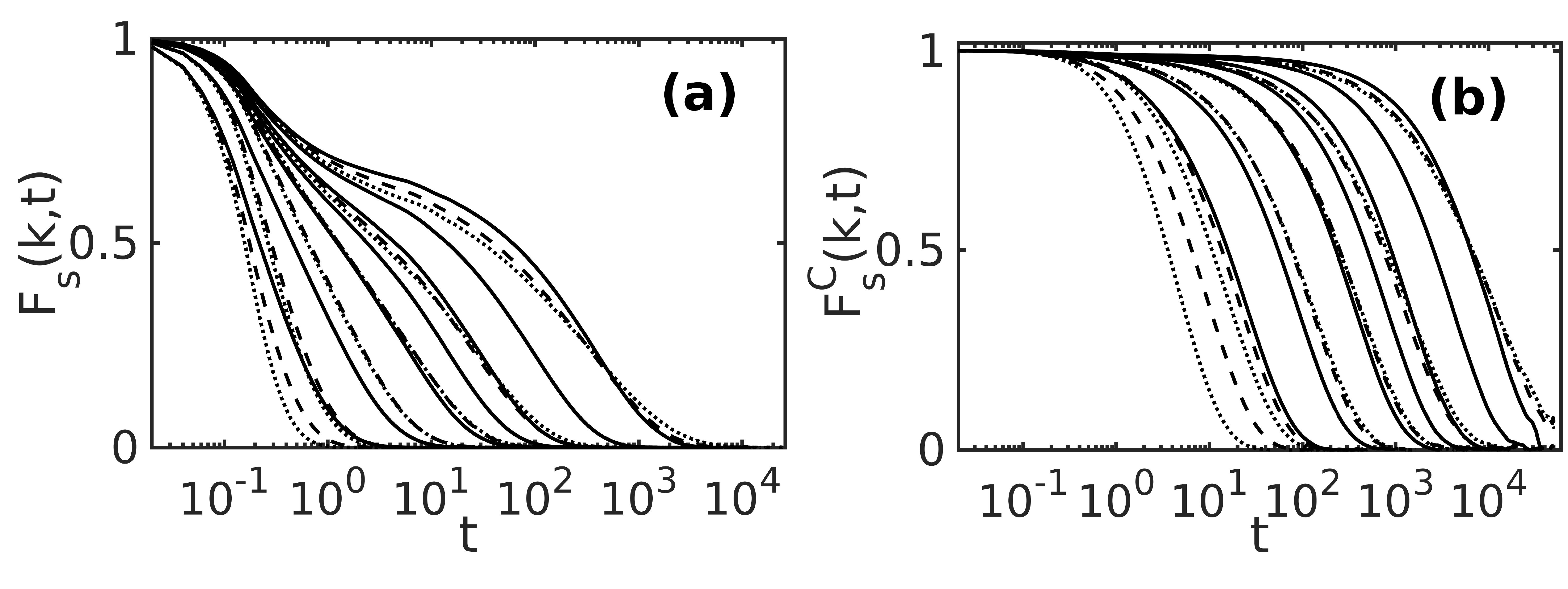}
	\caption{\label{f:dynpoly} Incoherent intermediate scattering function of monomers (a) and 
	Center of mass (b). Solid, dashed, and dotted lines correspond to the densities $\rho=$ 1.0, 
	$\rho=$ 0.85, and  $\rho=$ 0.7. In (a) and (b) temperature decreases from left to right 
	as $T=$ 2.0, 1.0, 0.6, 0.5, 0.45, 0.4, 0.36 ($\rho=$ 1.0 only).}
\end{figure}

Earlier studies argued that one of the possible reason for the violation of SE relation 
is the caging and hoping of the particles in supercooled liquids, which results
in the spatially heterogeneous 
dynamics \cite{j:edigermd,b:berthierdh,j:szamel2,j:SEparisi,s:chong_hop,j:sun_SE-SED_Janus}.
The slow down of density relaxation due to transient caging, and an increase in the
relaxation time with a reduction in the temperature, can be examined from the $F_s(k,t)$. 
Figure \ref{f:dynpoly}(a) shows $F_s(k,t)$ of monomers at the wavenumber $k=$ 6.9 for 
$\rho=$ 0.85, and $k=$ 7.1 for $\rho=$ 0.7 and $\rho=$ 1.0, at all the temperatures. As
temperature reduces, a hump appears in $F_s(k,t)$ from temperature $T=$ 0.5 for both lower 
densities, whereas it starts from $T=$ 0.6 for the higher density system. The appearance of 
a hump in the $F_s(k, t)$ of all three systems indicates a commencement of monomer caging that
enhances their $\alpha$-relaxation time. It is evident from Fig. \ref{f:dynpoly}(a) that
$F_s(k,t)$ of both lower density systems shows slower relaxation dynamics than that of the
higher density system at $T=$ 0.4, thus show a crossover. The heterogeneity in the relaxation 
is quantified from fitting the tail of $F_s(k,t)$ with empirical Kohlrausch-Williams-Watts (KWW) 
function $f(t)^{KWW}\!\propto\ \exp[-(t/\tau)^\beta]$ below temperature $T=$ 2.0 with the 
value of exponent $\beta$ varies from 0.92 to 0.59 for $\rho=$ 0.85, 0.94 to 0.55 for $\rho=$ 0.7,
and from 0.85 to 0.72 for the higher density system. This shows that lower densities 
exhibit more heterogeneous dynamics, which is a result of heterogeneous density distribution due
to the formation of the cavities (see nearest neighbour distribution in Fig. \ref{f:pnb}). We fit
$\tau_{\alpha}$ \textit{vs} $T$ curves (see Fig. \ref{f:talffit}) using schematic mode coupling 
theory (MCT) and Vogel-Fulcher-Tammann (VFT) relations, respectively, 
as $\tau_{\alpha}\!\propto (T-T_c)^{-\gamma}$ and $\tau_{\alpha}\!\propto \exp(AT_0/(T-T_0))$. 
The fitting parameters of these two equations for the density $\rho=$ 1.0, 0.85, and 0.7 are
shown in Table \ref{t:fit}. The fragility parameter $A$ obtained from the VFT fit shows that   
lower density systems are more fragile than the higher density system.
\begin{table}
	\caption{Fitting parameters for the MCT and VFT relations at three densities:
	MCT transition temperature $T_c$, MCT exponent $\gamma$, dynamic divergence temperature $T_0$,
	and fragility parameter $A$. $T_c$ and $T_0$ are lower for the higher density system and 
	fragility is smaller as fragility $\propto$ $1/A$ \cite{j:angel}. \label{t:fit}}	
	\begin{tabular}{|c|c|c|c|c|}
		\hline
		\multirow{2}{*}{\textbf{Density}} & \multicolumn{2}{c|}{\textbf{MCT fitting}} & \multicolumn{2}{c|}{\textbf{VFT fitting}} \\
		\cline{2-3} \cline{4-5}
		& $T_c$ & $\gamma$ & $T_0$ & $A$ \\
		\hline
		$\rho=$ 1.0 & 0.33 & 1.79 & 0.19 & 6.6 \\
		\hline
		$\rho=$ 0.85 & 0.39 & 1.71 & 0.31 & 2.12 \\
		\hline
		$\rho=$ 0.7 & 0.39 & 1.87 & 0.3 & 2.69 \\
		\hline
	\end{tabular}
\end{table}

The molecular density relaxation is studied from the time variation of molecular 
self-intermediate scattering function
\begin{equation}
	F^{C}_s(k,t)= N_c^{-1}\left\langle\rho^C_{\mathbf{k}}(t)\rho^C_{-\mathbf{k}}(0)\right\rangle,
\end{equation}
where $\rho^C_{\mathbf{k}}(t)=\exp[-i \mathbf{k} \cdot \mathbf{r}^{cm}(t)]$
and $\mathbf{r}^{cm}(t)$ is the position of the center of mass of a polymer molecule at time $t$.
Figure \ref{f:dynpoly}(b) shows that the $F^{C}_s(k,t)$ of lower and higher density
systems are plotted against time $t$ at $T=$ 2.0-0.36, and the wavenumber
$k=2\pi/2R_g\sim $ 2.1. Here, $R_g$ is the average radius of gyration of polymer chains in 
both lower and the higher density systems, and its values are 1.46 and 1.45. Thus, molecular 
diameters are $2R_g=$ 2.92 and 2.9, and we consider $2R_g=$ 2.9. 
In Fig. \ref{f:dynpoly}(b), the $F^C_s(k,t)$ of both lower density systems shows
crossover to the longer relaxation time in comparison to the higher density system at the same 
temperature $T=$ 0.4; the $F^C_s(k,t)$ of both lower density systems at $T=$ 0.4 is comparable 
to the $F^C_s(k,t)$ of the higher density system at $T=$ 0.36. To examine the time scale of the 
molecular relaxations, we calculate molecular relaxation 
time $\tau_{2R_g} = \int_0^{\infty} F^C_s(k,t) dt$, which is higher at high (and intermediate) 
temperatures for the higher density system compare to both lower density systems. However, in
the moderately supercooled regime (near $T_c$) there is a crossover similar to the case of monomer
density relaxation (see Fig. \ref{f:orlxtym}). Now, we look into the violation of SE and SED
relations at these state points of the system.

\subsection{\label{se:sev} Violation of Stokes-Einstein relations}
\begin{figure}
	\includegraphics[width=8.5cm, height=6.5cm]{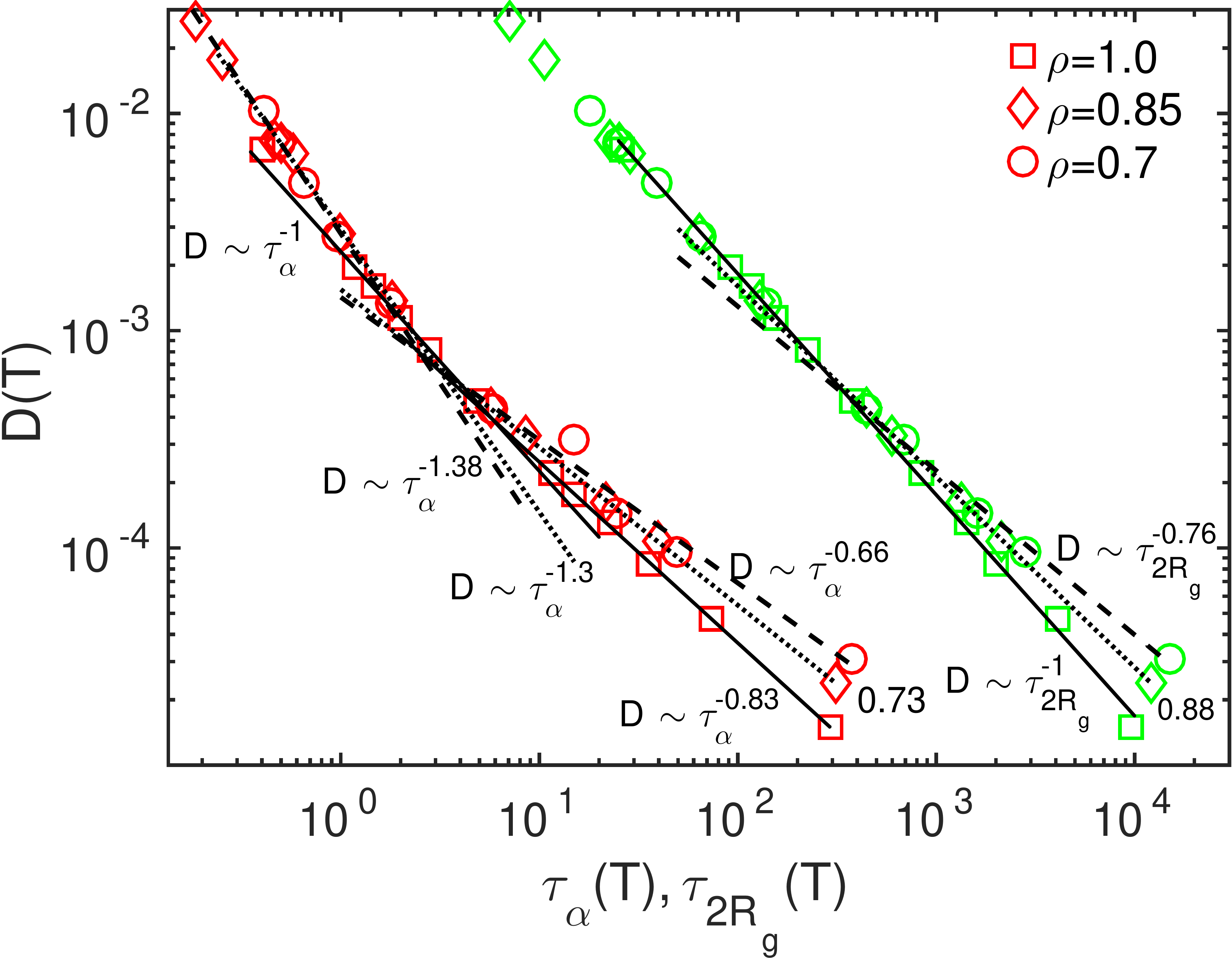}
	\caption{\label{f:sevp} Self diffusion coefficient $D(T)$ is plotted against
	$\tau_\alpha(T)$ (red) and $\tau_{2R_g}(T)$ (green). Solid, dotted, and dashed lines are fit
	to the data at densities $\rho=$ 1.0, 0.85, and 0.7, respectively: the data is fitted using
	relations $D \sim \tau^{-\xi}_\alpha$ and $D \sim \tau^{-\xi}_{2R_g}$ for monomers and
	molecules, respectively.}
\end{figure}

To look at the violation of SE relation, we compute exponent of the
power-law dependence of diffusion constant on the relaxation time, i.e., 
$D\sim\tau^{-\xi}_\alpha$ at different densities. Figure \ref{f:sevp} shows that
$\xi=$ 1.0 for the higher density, whereas $\xi=$ 1.38 and 1.3 for $\rho=$ 0.7 
and 0.85, respectively, in the temperature range $T=$ 2.0--0.6.  Exponent $\xi>1$ shows
the decoupling of $D$ and $\tau_\alpha$ in the normal liquid temperatures, which is also
reported in many studies including Refs. \cite{j:shiladitya,j:ganapathy}. In moderately 
supercooled regime ($T=$ 0.5--0.36), the value of $\xi=$ 0.83 for the higher density system, 
and $\xi=$ 0.73 and 0.66 for $\rho=$ 0.85 and 0.7, respectively, thus shows decoupling of $D$ 
and $\tau_\alpha$. The decreasing value of $\xi$ with the density suggests that the decoupling
increases with decreasing density. A smaller value of $\xi$ in the supercooled regime of both 
lower density systems shows the pronounced violation of the SE relation, and in particular,
the violation is more pronounced for the lower density $\rho=$ 0.7. In both lower density 
systems, effective attraction between particles is pronounced, in contrast to the higher
density system. A study on attractive and repulsive colloidal glasses also shows the 
stronger violation of the Stokes-Einstein relation in the attractive glassy 
colloids \cite{attrColSE} .

\begin{figure}
	\includegraphics[width=8.5cm, height=6.5cm]{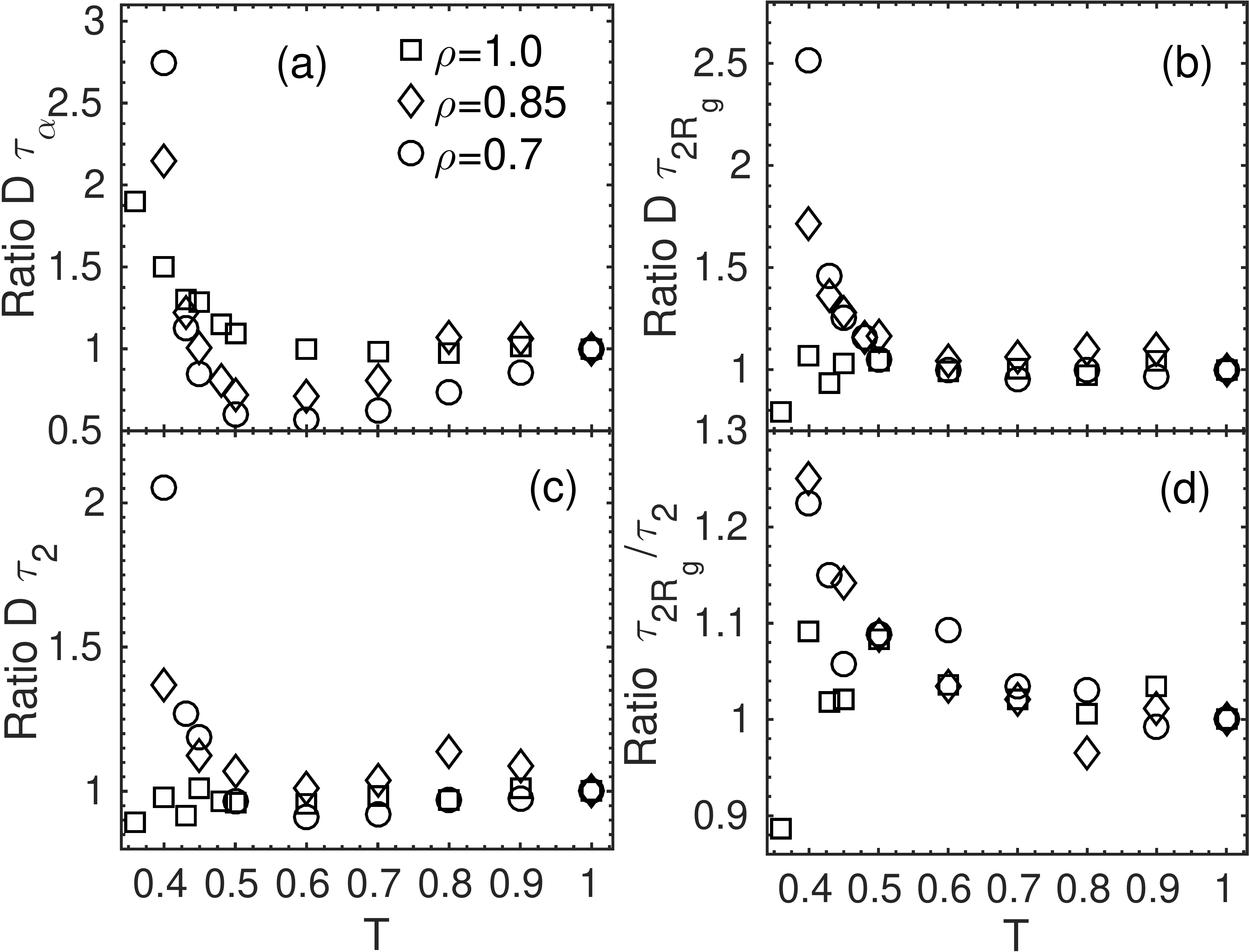}
	\caption{\label{f:sedvp1} Ratios of (a) $D\tau_\alpha$, (b) $D{\tau_{2R_g}}$,
	(c) $D\tau_2$, and (d) $\tau_{2R_g}/\tau_2$ are plotted against temperature $T$.
	The ratios are defined as $D\tau(T)/D\tau(T=1.0)$, where $D\tau(T=1.0)$ is a reference 
	point. }
\end{figure}

Another way of estimating the SE breakdown is the predictors of violation of the
SE relation \cite{j:shiladitya,j:bhowmik_sk,s:deb_stil,j:onuki}, e.g., 
$D\tau_\alpha(T)$. We compute SE ratio as $D\tau_\alpha(T)/D\tau_\alpha(T=1.0)$,
where $D\tau_\alpha$ at temperature $T=$ 1.0 is a reference point, as given in 
Fig. \ref{f:sedvp1}(a). Figure \ref{f:sedvp1}(a) shows that the SE ratio starts
decreasing below $T=$ 1.0, which again starts increasing from $T=$ 0.5 and reaches values 
around 2.1 and 2.7, respectively, at the lowest temperature $T=$ 0.4 of the density
$\rho=$ 0.85 and $\rho=$ 0.7. In the higher density system, the SE ratio remains constant up 
to $T=$ 0.6, which starts increasing from $T=$ 0.5 and attains a value around 2.0 at the lowest 
temperature $T=$ 0.36. Below temperature $T=$ 0.5, where fractional SE relation is found at all
three densities (see Fig . \ref{f:sevp}), the SE ratio also starts increasing. A higher value
of the SE ratio  correlates with the smaller value of the fractional SE exponent $\xi$.
Thus, we show that the SE relation breaks down in the supercooled linear polymers for
all three densities.

Polymer chains in this study are flexible, which shows large shape fluctuations. Therefore,
it is compelling to examine the violations of the SE relation at the molecular level. Diffusion 
constant $D$ is plotted against molecular relaxation in Fig. \ref{f:sevp}, where data is fitted
using the relation $D \sim \tau^{-\xi}_{2R_g}$. A variation in $\tau_{2R_g}$ with temperature
is shown in Fig. \ref{f:orlxtym}. The value of the exponent $\xi$ is 1.0 from $T=$ 2.0--0.6 for
both lower density systems. However, in the supercooled regime ($T=$ 0.5--0.4), the value of 
the exponent $\xi$ are 0.88 and 0.76, respectively for the density $\rho=$ 0.85 and 
$\rho=$ 0.7, thus, show a weak (exponent $\xi$ closer to unity) violation of molecular SE 
relation; the extent of violation is more  at $\rho=$ 0.7. On the other hand, 
$\xi=$ 1.0 for whole temperature range (of this study) at the higher density,
thus, $D$ and $\tau_{2R_g}$ are coupled. We examine the SE  ratio of
$D\tau_{2R_g}$ also in all three systems and found that in the temperature range where
$D \sim \tau^{-\xi}_{2R_g}$ is fractional, the SE ratio also starts increasing from the value 1.0 
and reaches a  value around 1.65 and 2.5 respectively for $\rho=$ 0.85 and $\rho=$ 0.7
[see Fig. \ref{f:sedvp1}(b)]. However, in the higher density system, this ratio oscillates 
around 1.0. Thus, this study shows the (weak) violation of the molecular SE relation only in
the lower density systems, whereas the violation of monomer SE relation is seen at all 
three densities. Many investigations show the role of mobile and immobile species and jump like
motions in the violations of the SE relation, therefore, next, we look for many possible
origins of violations of SE relations in this flexible unentangled polymer system.

\subsection{Particles' mobility and jumplike motion \label{se:mob}}

\begin{figure}
	\includegraphics[width=7.0cm,height=5.0cm]{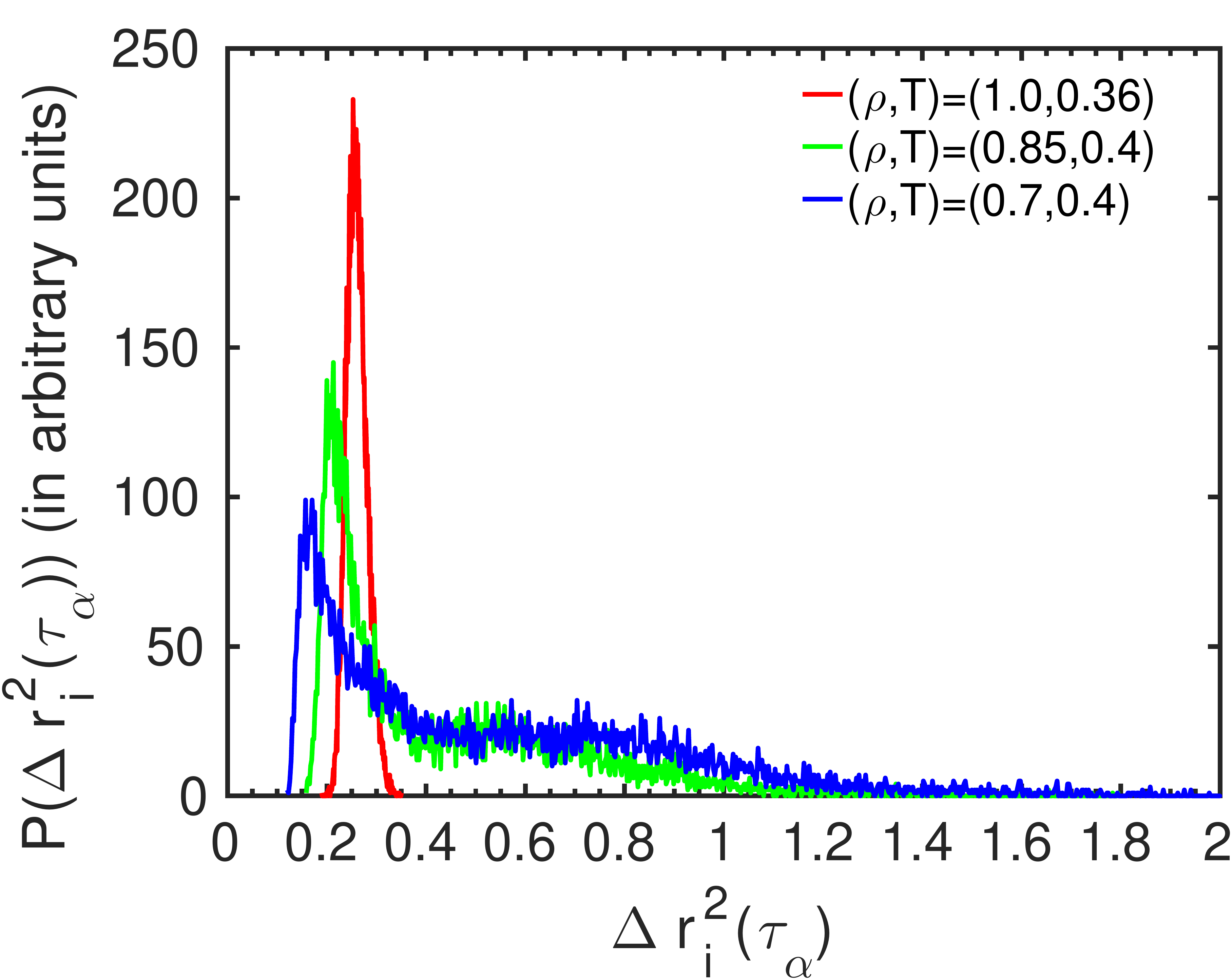}
	\caption{\label{f:pmsdpp} Probability distribution of squared displacements
	of individual monomers, $P[{\Delta r_i}^{2}(\tau_\alpha)]$, averaged over
	time difference $t=\tau_\alpha$, at the lowest temperatures of the study where
	violation of the SE relation is pronounced.}
\end{figure}

\begin{figure}
	\includegraphics[width=8.3cm, height=3.8cm]{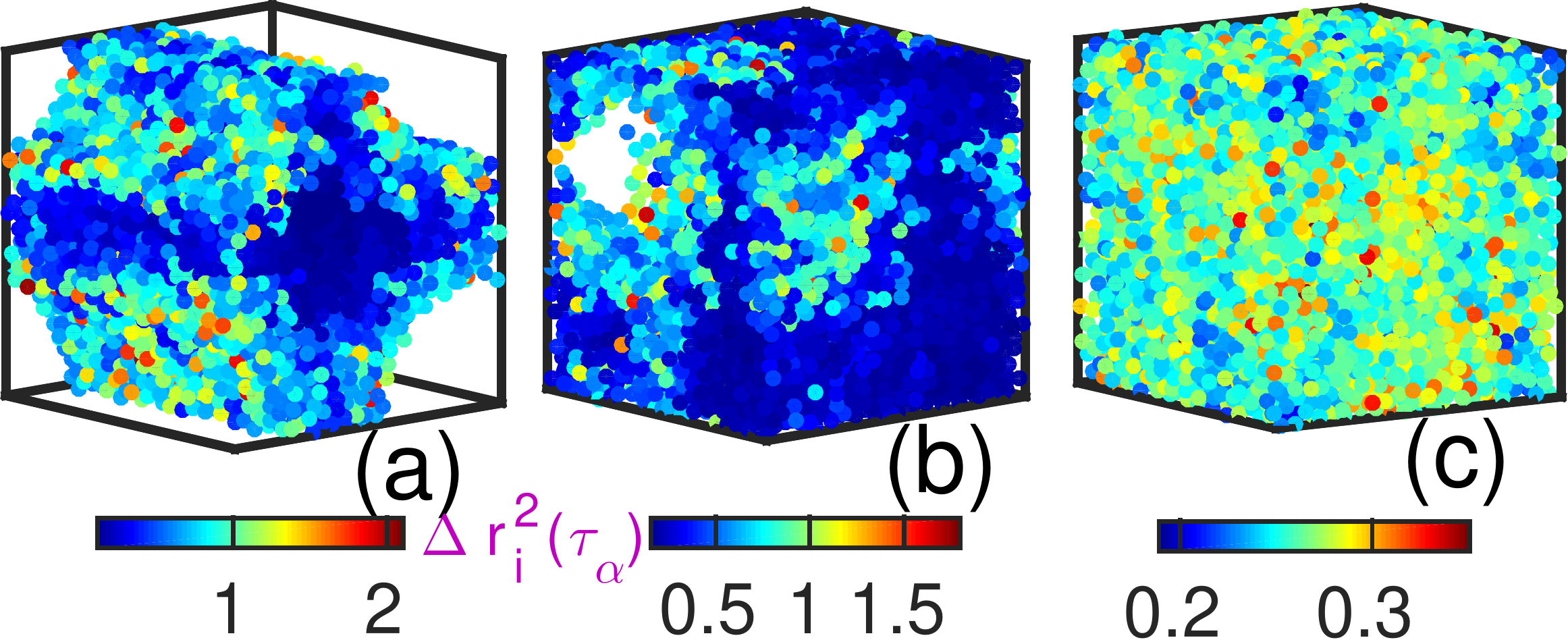}
	\caption{\label{f:msdmap} Squared displacement of each monomer at
	$t=\tau_\alpha$, i.e., $\Delta r_i^2(\tau_\alpha)$ of (a) density $\rho=$ 0.7 
	at $T=$ 0.4, (b) density $\rho=$ 0.85 at $T=$ 0.4, and (c) density $\rho=$ 1.0 
	at $T=$ 0.36. Color bars are according to $\Delta r_i^2(\tau_\alpha)$.}
\end{figure}

To examine the mobility of the particles in the system, we compute squared 
displacement of each particle (monomer), i.e., 
\begin{equation}
	\label{e:sd}
	{\Delta r_i}^{2}(\tau_\alpha)=\langle|\mathbf{r}_i(\tau_\alpha)-\mathbf{r}_i(0)|^{2}\rangle,
\end{equation}
which is averaged over the time difference, $t=\tau_\alpha$, not over the particles. The 
probability distribution of ${\Delta r_i}^{2}(\tau_\alpha)$, displayed in
Fig. \ref{f:pmsdpp}, shows a distribution of squared displacements at 
time $t=\tau_\alpha$ (average cage-relaxation time of monomers). 
Figure \ref{f:pmsdpp} shows that ranges of ${\Delta r_i}^{2}(\tau_\alpha)$
are 0.13--2.1 and 0.16--1.7, respectively for $\rho=$ 0.7 and 0.85, at $T=$ 0.4, 
whereas ${\Delta r_i}^{2}(\tau_\alpha)$ varies as 0.19--0.35 for the 
higher density at $T=$ 0.36, showing a significant difference in the range of 
$P[{\Delta r_i}^{2}(\tau_\alpha)]$. Interestingly, $P[{\Delta r_i}^{2}(\tau_\alpha)]$ of
both lower density systems shows a hump (in addition to the main peak), which corresponds
to the monomers that are faster than the average motion of the monomers in the system.
Thus, it shows a disparity in ${\Delta r_i}^{2}(\tau_\alpha)$ of monomers in both lower
density systems compared to the higher density system where the range
of ${\Delta r_i}^{2}(\tau_\alpha)$ is narrow and no hump is present. For getting the spacial
distribution of square displacements, we show the configurations of the 
systems with their squared displacements as a color map in Figs. \ref{f:msdmap}(a) 
and \ref{f:msdmap}(b), where both lower density systems show macroscopic cavities, however,
the higher density system shows a continuous density distribution across the 
system [see Figs. \ref{f:pnb}(a--c) for average nearest-neighbor distributions]. In both
lower density systems at $T=$ 0.4, the monomers near the cavities show larger squared
displacements at $t=\tau_\alpha$, whereas the monomers in the core show smaller squared 
displacements, i.e., the monomers in the core start freezing, whereas the monomers near the
cavities are in the gaseous phase, which creates a disparity in their squared displacements
resulting in the pronounced dynamic heterogeneity \cite{j:ls_paddy4} in both lower density 
systems. The dynamical heterogeneity is less pronounced in $\rho=$ 0.85 system than 
the $\rho=$ 0.7 system because of a bit narrower distribution of the squared displacements
at $t=\tau_\alpha$, as the surface of the cavities is reduced. This analysis suggests that an
extent of the dynamical heterogeneity in the monomer motion of both lower densities arises due
to a large disparity in their displacements because of the presence of surfaces around the 
cavities \cite{jalim:ppj2}. As $P[{\Delta r_i}^{2}(\tau_\alpha)]$ shows disparity in the
distribution, we look at more averaged probability distributions, later in this study,
thus attempt to correlate the SE violation with the observed disparity in the mobility.
Now, it is interesting to look at the characterization of dynamic cages from the analysis
of single particle motion, averaged over time differences and number of particles.

\begin{figure}
	\includegraphics[width=6.5cm,height=8.5cm]{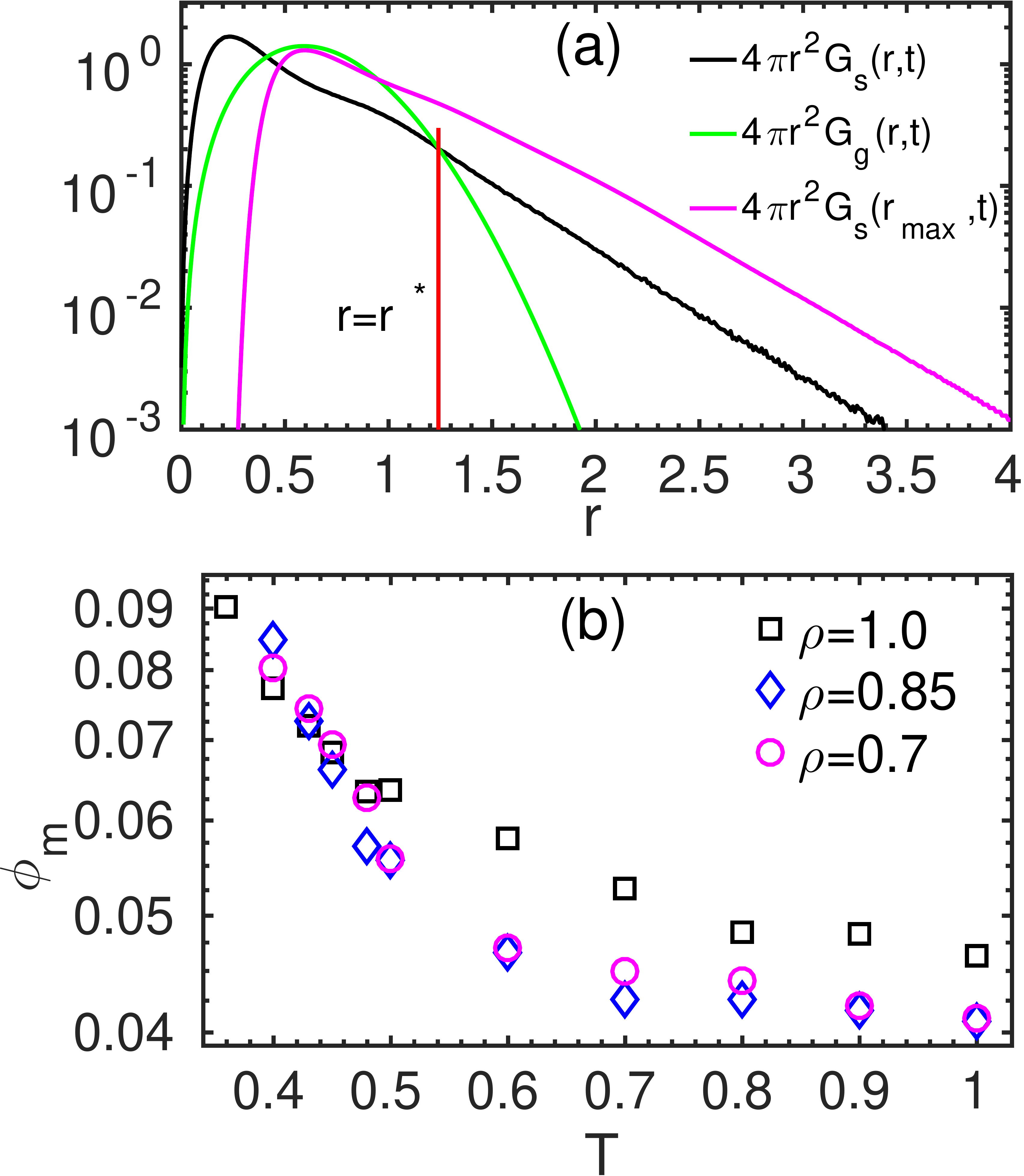}
	\caption{\label{f:gaus} van Hove correlation function $G_s(r,t)$, its Gaussian
	approximation $G_g(r,t)$, and probability of maximum displacements $G_s(r_{max},t)$, 
	are compared at time $t=\tau_\alpha$ for $\rho=$ 0.7 at $T=$ 0.4 (a). A variation in
	a fraction of mobile particles with temperature at densities $\rho=$ 1.0, 0.85, 
	and 0.7 (b). Mobile particles are characterized as the particles that are moved greater
	than the distance $r^*$ at time $t$.}
\end{figure}

Jumplike motion of the particles in the system can be identified by self part of 
van Hove correlation 
function \cite{ham,wahnstrom,sastry_jump,expt_colloid,kawasaki_jump,jump_lam},
\begin{equation}
	\label{e:gsrt}
	G_{s}(\mathbf{r},t) = \frac{1}{N} \left\langle {\sum\limits_{i=1}^{N}
	{\delta[\mathbf{{r}}+{\mathbf{r}}_{i}(0)-{\mathbf{r}}_{i}(t)]}} \right\rangle.
\end{equation}
The long tail in the $G_s(r,t)$ corresponds to the particles having larger displacements,
thus, higher mobility. To calculate a cutoff radius $r$ and a fraction of these higher 
mobility particles, we compare the $G_s(r,t)$ with its Gaussian approximation at the
same time, i.e., $\tau_\alpha$
\begin{equation}
	\label{e:ggrt}
	G_g(r,\tau_\alpha) = \left[ \frac{3}{2\pi\langle r^2(\tau_\alpha)\rangle}\right]^{3/2}
	\exp\left[ \frac{-3 r^2}{2\langle r^2(\tau_\alpha)\rangle}\right] ,
\end{equation}
which is displayed in Fig. \ref{f:gaus}(a). Here, $\langle r^2(\tau_\alpha)\rangle$ is an 
average MSD of the monomers in the time interval [t, t+$\tau_\alpha$].
$G_s(r,t)$ is Gaussian in the ballistic time regime and in the diffusive time regime. Between
these two time scales, it deviates from its Gaussian approximation given in Eq. \ref{e:ggrt}.
For comparison of $G_s(r,t)$ and $G_g(r,t)$, we choose to show (for example) the lower 
density $\rho=$ 0.7 at temperature $T=$ 0.4, which is shown in Fig. \ref{f:gaus}(a). We consider
a particle as mobile if it travels a distance equal to or greater than $r^*$. The distance 
$r^*$ is marked by the red line in Fig. \ref{f:gaus}(a), where $G_s(r,t)$ crosses over to
the $G_g(r,t)$. Thus, the fraction of mobile particles is computed as 
\begin{equation}
	\label{e:fracmob}
	\phi_m = \int_{r^*}^\infty 4\pi r^2 G_s(r^*,\tau_\alpha) dr ,
\end{equation}
as for e.g. in the binary LJ mixture \cite{j:scg4,j:scg3}. In the previous studies of glass 
transition properties on the binary LJ mixture \cite{j:scg4,j:scg3}  and 
polymers \cite{j:fwstarr}, the fraction of mobile particles is calculated at a time 
scale corresponding to a peak value of the non-Gaussian parameter. Here, we choose the time
scale as $t=\tau_\alpha$ for the calculation of a fraction of mobile particles because 
$\tau_\alpha$ is dominated by caged particles. However, the diffusion coefficient is
dominated by mobile particles. Therefore, the estimation of mobile particles' fraction
at time $t=\tau_\alpha$ can give information about the decoupling between $D$ and
$\tau_\alpha$. Figure \ref{f:gaus}(b) shows a variation in the fraction of mobile
particles, $\phi_m$, with $T$ and $\rho$ : $\phi_m$ increases with $T$ from 4\% to 9\%, 
at each density. In the temperature range $T=$ 1.0--0.6, $\phi_m$ is around
$\approx$ 4\% and 5\% for both lower densities and the higher density,
respectively. However, in the supercooled regime ($T=$ 0.5--0.36), $\phi_m$ increases 
rapidly with a decrease in $T$, for all three densities, though it does not show a systematic
variation with density. In Sec. \ref{se:sev}, we show that the SE violation increases with
a decrease in the density at the same $T$. Thus, $\phi_m$ does not correlate with the extent
of SE violations and the density of the system. This prompts us to look for a distribution
function of the maximum displacement of the particles.
\begin{figure}
	\includegraphics[width=7.3cm,height=11.3cm]{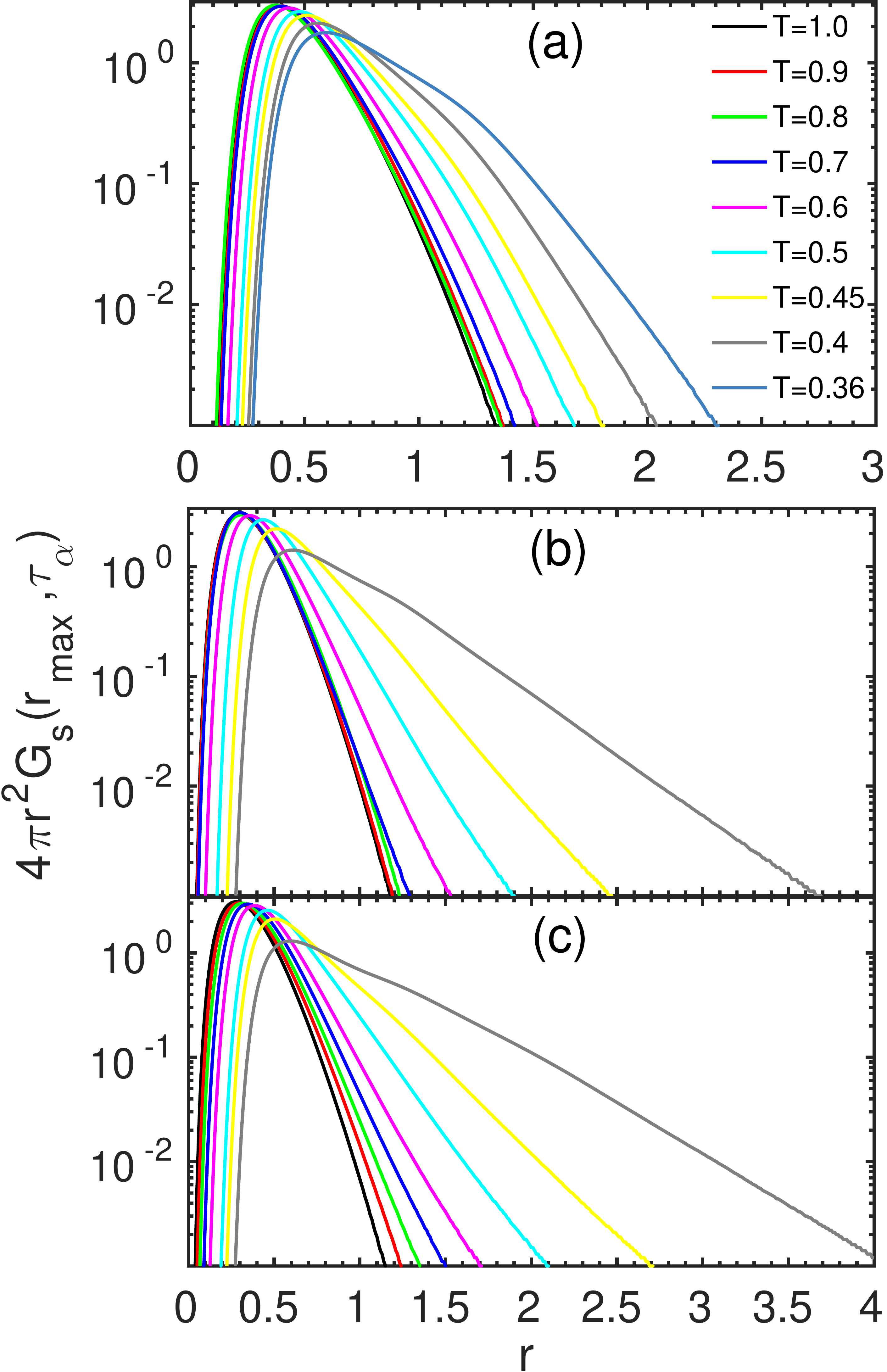}
	\caption{\label{f:gsrmaxt}  $4 \pi r^2 G_s(r_{max},\tau_\alpha)$ of monomers is plotted 
	against radial distance $r$ at densities (a) $\rho=$ 1.0, (b) $\rho=$ 0.85,
	and (c) $\rho=$ 0.7. The legend of (a) is applicable to (b) and (c).}
\end{figure}

As particles execute random walk (at the time scale of molecular diffusion) they
revisit their original position, thus, the maximum distance a particle travels during a 
time interval $t$ is not identified in the $G_s(r,t)$. To identify the distribution
of maximum displacement, we compute maximum displacement of a particle within a time
interval [t, t+t$'$], and then calculate its probability distribution similar to
the $G_s(r,t)$. We use the definition of maximum displacement given in 
Ref. \cite{j:scg3},
\begin{equation}
	r_{max}(t,\tau_\alpha) = \textnormal{max} \{ |\mathbf r(t+t^\prime) - \mathbf r(t)| \},
\end{equation}
where $t^\prime \in [0, \tau_\alpha]$. A probability distribution
of $r_{max}(t,\tau_\alpha)$, i.e., $4 \pi r^2 G_s(r_{max},\tau_\alpha)$ is calculated and
displayed in Figs. \ref{f:gsrmaxt}(a--c) to look at a variation in the mobility of 
particles with $T$ and $\rho$. This definition of mobility captures both type of 
particles, i.e., mobile particles, and most and least immobile particles. The long
tail of $4 \pi r^2 G_s(r_{max},\tau_\alpha)$ corresponds to the mobile particles. 
A systematic spread in $4 \pi r^2 G_s(r_{max},\tau_\alpha)$ with temperature
shows a difference in different mobilities of particles for all three densities, which
measures the disparity in particles' displacements. In the supercooled regime, the 
spread in the tail of $4 \pi r^2 G_s(r_{max},\tau_\alpha)$ is more in both lower
density systems compared to the higher density system, showing more disparity in the 
particles' displacements. In Sec. \ref{se:sev}, we have shown that the SE violations 
are more pronounced in both lower density systems compared to the higher density 
system at the same $T$. Thus, the disparity in the particles' motion is highly correlated 
with the violation of the SE relation in these linear polymer chains, instead of only the 
fraction of mobile particles. Furthermore, we show that the SE violations vary with the 
system size, which is related to the disparity in the particles' 
mobility (see \ref{sa:finsize} for a detailed explanation). Next, we present the caging 
and jumplike motion of the particles, directly from their translational trajectories.


\begin{figure}
	\includegraphics[width=7.5cm, height=7.5cm]{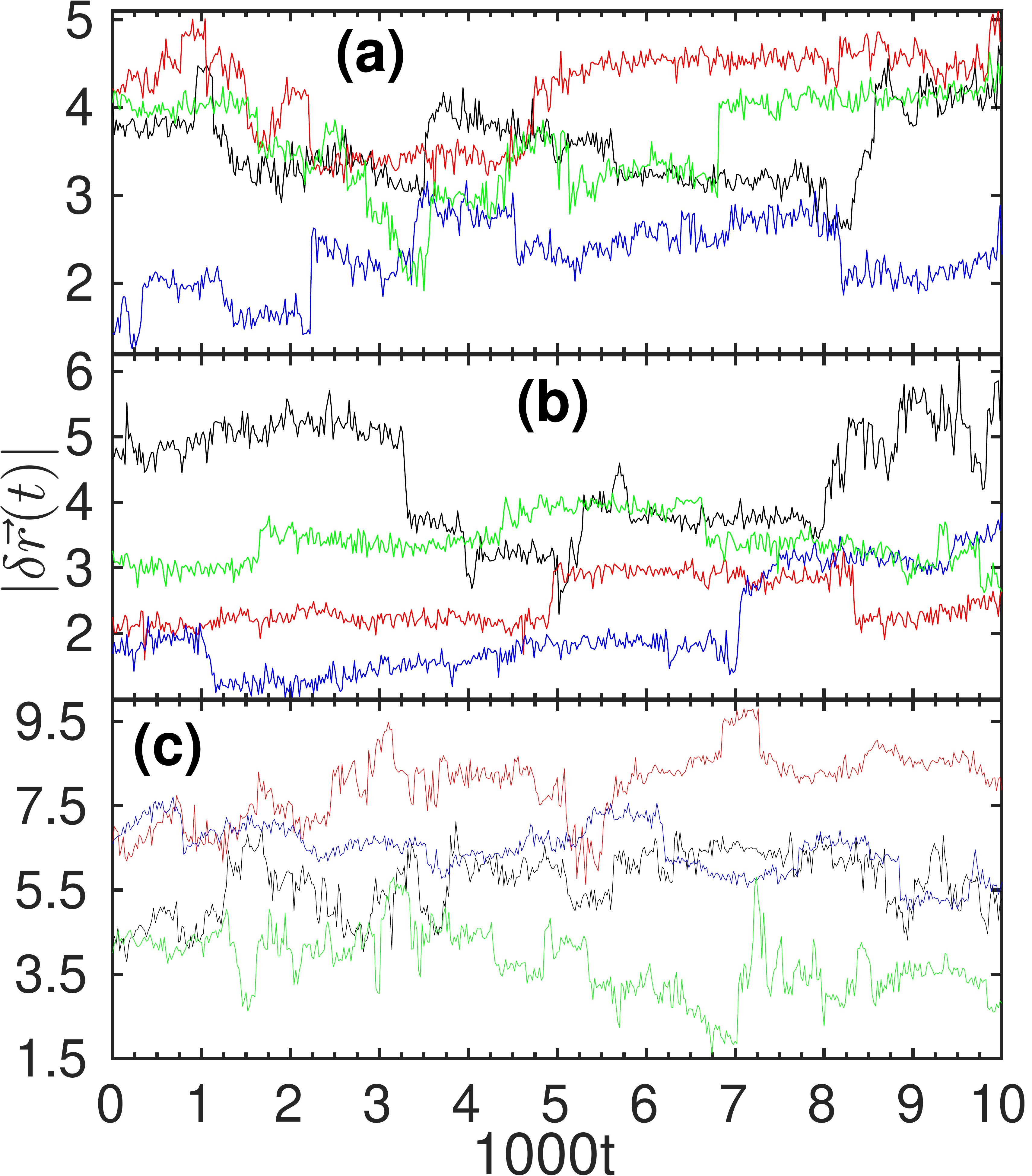}
	\caption{\label{f:trtrj} Trajectories of the monomer displacements,
	$|\delta \vec r(t)|=|\vec r_i(t)-\vec r_i(0)|$ for (a) $\rho=$ 1.0 at $T=$ 0.36, 
	(b) $\rho=$ 0.85 at $T=$ 0.4, (c) $\rho=$ 0.7 at $T=$ 0.4. Different colors in 
	(a), (b), and (c) correspond to the randomly selected monomers of different polymer
	chains.}
\end{figure}

In the supercooled regime, due to the presence of molecular cages, these molecules
undergo large directed displacements by translation to escape from the self-generated 
barriers, thus relaxing the accumulation of the stress due to hindered motion. To identify
such motions, we plot typical trajectories of the displacements of the randomly
selected single monomers $|\mathbf{r}_i(t) - \mathbf{r}_i(0)|$  against time $t$ at the lowest 
temperatures $T=$ 0.4 and 0.36 to see whether there are large displacements in the
monomers. Figures \ref{f:trtrj}(a--c) show the typical trajectories of $\rho=$ 0.7 and 0.85 system
at $T=$ 0.4, and $\rho=$ 1.0 system at $T=$ 0.36. These trajectories show the intermittent large
displacements in the motion of monomers that have deviations from the regular random walk.  
Jumplike motions are difficult to identify in polymer systems with longer chains. An earlier 
study using continuous-time random walk on shorter chains ($n=$ 4) identifies jumps in the
supercooled linear polymer melt  \cite{j:helfrich1}. However, jump like motions are found in
several studies of glass-forming binary LJ systems (see
Refs. \cite{j:lberthier,j:sarikahop}). These intermittent large displacements of the monomers are
related to the fluctuations in the molecular configurations, which appear at the same state points
where the violation of the SE relation is much pronounced, i.e., at $T=$ 0.4 and 0.36 for 
the lower density and higher density systems, respectively. 


\subsection{\label{se:ori} Violation of Stokes-Einstein-Debye relations}

\begin{figure}
	\includegraphics[width=8.0cm, height=6.0cm]{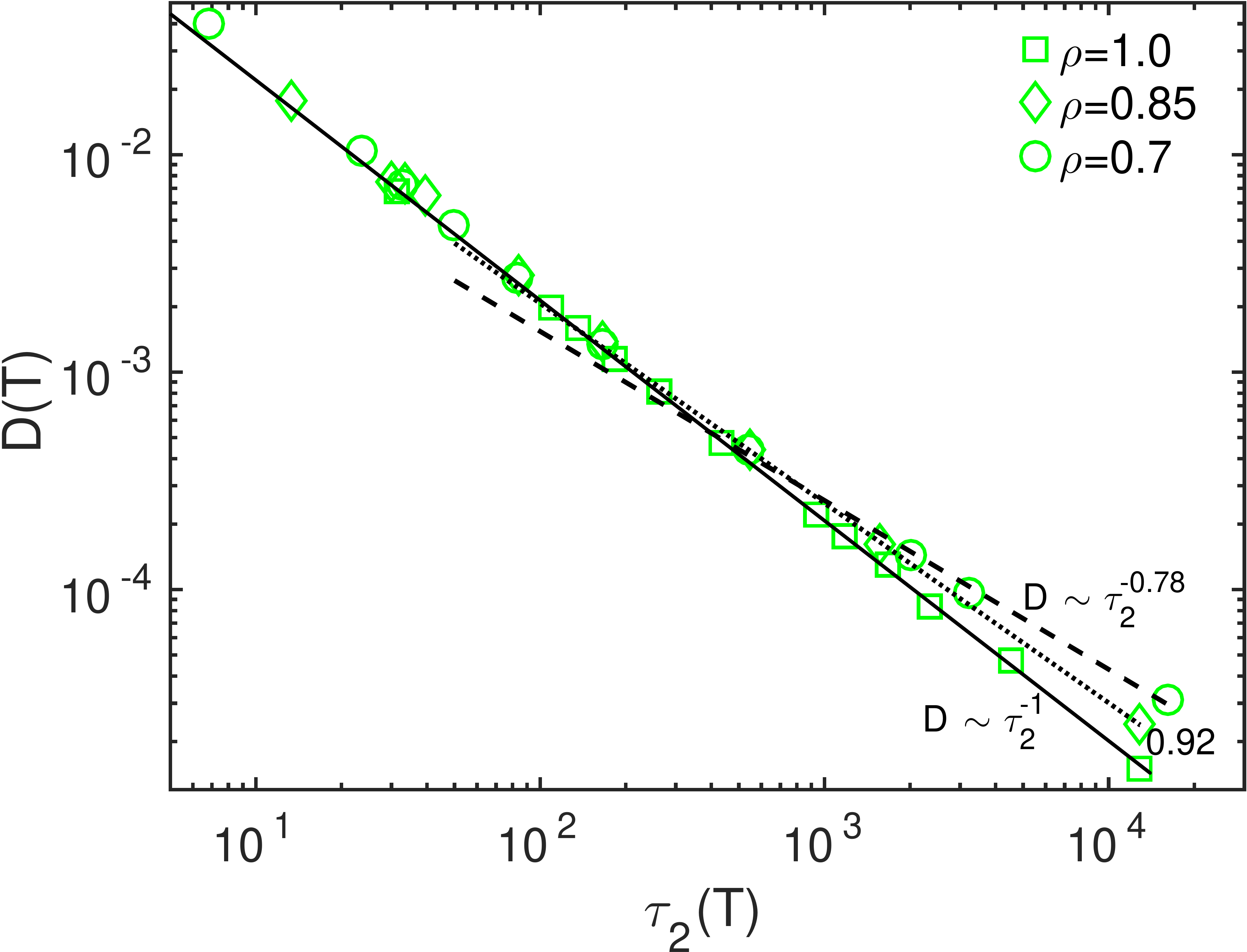}
	\caption{\label{f:sedvp} Diffusion coefficient $D(T)$ 
	is plotted against $\tau_2(T)$. Solid, dotted, and dashed lines are fit to the data at
	densities $\rho=$ 1.0, 0.85, and 0.7, respectively; the data is fitted using relation
	$D\sim \tau^{-\xi}_2$.}
\end{figure}

Polymers are extended macromolecules that have rotational degrees of freedom along with
translational ones. The average rotational motion of polymer molecules can be quantified 
from the rotation of the unit vector along the end-to-end vector. The relaxation time of the 
rotational motion is calculated as $\tau_l =\int_0^\infty C_l^{r}(t)dt $,
where $C_l^r(t) = \langle{P_l}[\cos\theta(t)]\rangle$
is $l^{th}$ order rotational correlation function for a non-spherical molecules \cite{bap}. 
Here, $\cos\theta(t) = \mathbf{\hat e}(0).\mathbf{\hat e}(t)$, and $\mathbf{\hat e}(t)$ is an 
unit vector along an end-to-end vector $\mathbf e(t) = \mathbf r_{1}(t) - \mathbf r_{n}(t)$
of a polymer chain; ${P_l}[\cos\theta(t)]$ is the Legendre polynomial of order $l$. 
Molecular liquids show translation-rotation coupling at high temperatures and obey the 
relation $D\tau_l=$ constant, which means that $D\propto\tau_l^{-1}$. Many studies show that
near the glass transition the translation-rotation relaxation dynamics is 
decoupled \cite{s:chong_kob,j:stilhodg}. The failure of SED relation is usual in the 
molecular liquids \cite{j:dsefail}, which is much more pronounced in the orientatinal glasses,
e.g., see Refs. \cite{s:chong_kob,bb:ppj2}. The polymer model, we have used in this study, is a
non-polar, therefore, we have computed orientational relaxation time, especially, $\tau_2$
due to up-down symmetry, to examine the SED violation \cite{j:leporini5,j:dse_exp}. Recently,
dependency of SED relation on the degree of Legendre polynomial, $l$, is examined by
Kawasaki and Kim in supercooled water \cite{j:kim_kawasaki_ldep}. Data of $D$ \textit{vs}
$\tau_2$ is fitted with the relation $D \sim \tau^{-\xi}_2$, as shown in Fig. \ref{f:sedvp},
and the variation of $\tau_2$ with temperature is shown in Fig. \ref{f:orlxtym}.
In both lower density systems, exponent $\xi=$ 1.0 from $T=$ 2.0--0.6, whereas in the 
supercooled regime ($T=$ 0.5--0.4), the exponent $\xi=$ 0.92 and $\xi=$ 0.78, respectively
for $\rho=$ 0.85 and $\rho=$ 0.7 system. These fractional values of $\xi$ show a weak 
decoupling in $D$ and $\tau_2$ near $T_c$ in both lower density systems, which is significant
in the $\rho=$ 0.7 system. On the other hand, in the higher density system, $\xi=$ 1.0 from
high to low temperatures (up to $T=$ 0.36), which means that $D$ and $\tau_2$ are coupled 
even near the $T_c$.
\begin{figure}
	\includegraphics[width=7.0cm,height=5.0cm]{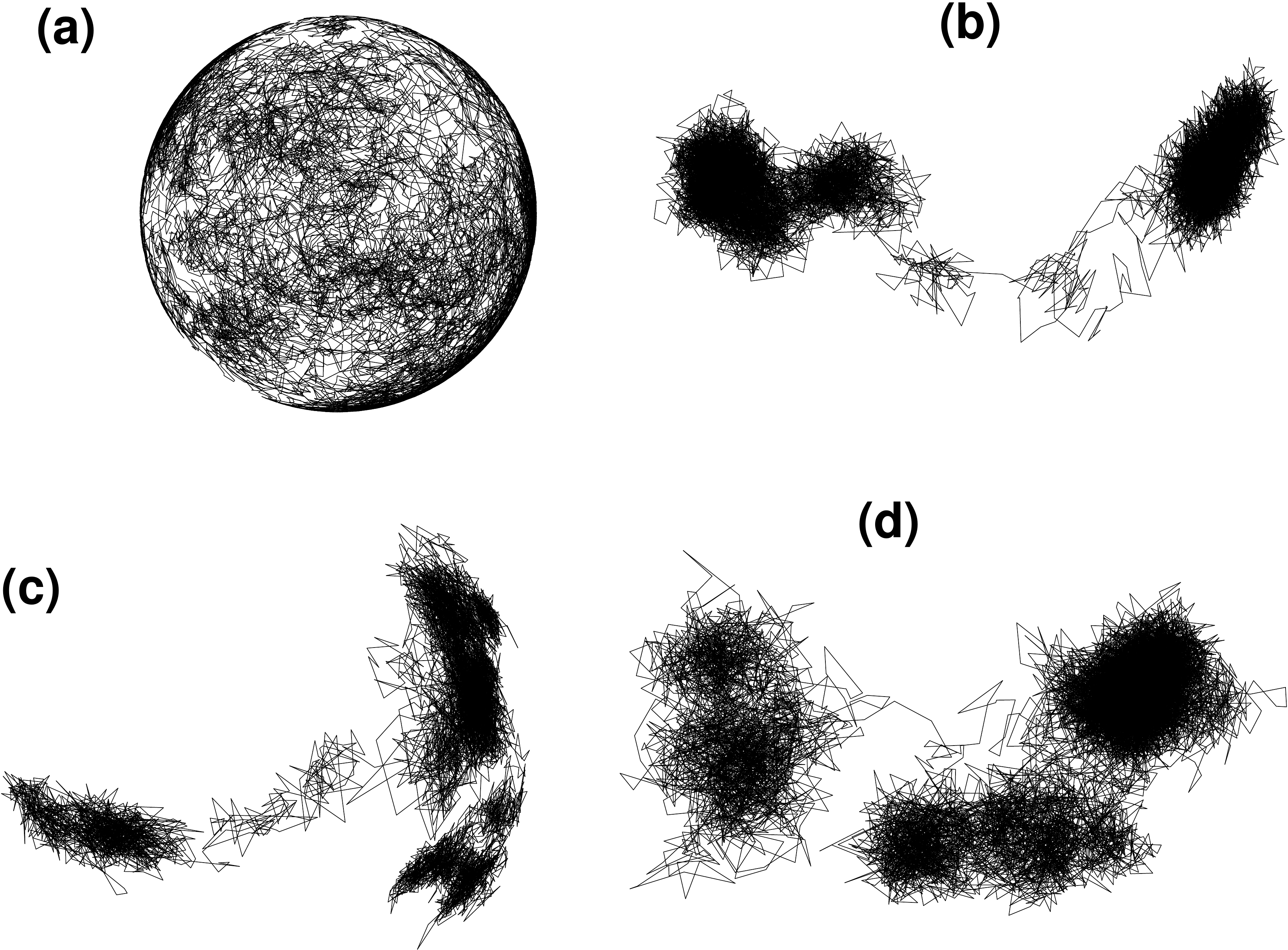}
	\caption{\label{f:or} The trajectory of unit vector along the end-to-end vector
	of a randomly selected chain of the system for the time interval $\delta t=$ 2500.
	(a) $\rho=$ 0.85 at $T=$ 1.0, (b) $\rho=$ 0.85 at $T$ = 0.4, (c) $\rho=$ 1.0 at 
	$T$ = 0.4, and (d) $\rho=$ 1.0 at $T$ = 0.36.}
\end{figure}
Analogous to the ratios of predictors of SE relation, we compute ratio of $D\tau_2=$ constant
as a predictor for the SED violation \cite{j:tarjus} . The value of $D\tau_2$ is scaled with
its value at $T=$ 1.0, which is shown in Fig. \ref{f:sedvp1}(c) (ratio is defined 
as $D\tau_2(T)/D\tau_2(T=1.0)$). In both lower density systems, this ratio starts from 1.0 and
fluctuates around this value from $T=$ 0.9 to $T=$ 0.6. It starts increasing from $T=$ 0.5 and
reaches around values 1.3 and 2.0 (at $T=$ 0.4) for $\rho=$ 0.85 and $\rho=$ 0.7 systems, 
respectively. This shows that SED relation is weakly violated in $\rho=$ 0.85 system, similar
to the molecular SE violations, whereas violation is more for the $\rho=$ 0.7 system. In the
temperature range where there is a maximum value of SED ratio, the exponent $\xi$ reduces,
thus shows a correlation between them. In the higher density system, the SED ratio fluctuate
close to 1.0 except at the lowest temperature $T=$ 0.36, where the ratios attains a value around
0.8. This implies that SED relation is valid in the higher density system even near $T_c$.
An experimental and theoretical study by Gupta \textit{et al.} shows that the 
Stokes-Einstein relation obeys up to the glass transition in starlike micelles owing to 
ultrasoft interactions \cite{j:se_richter_valid} . Further, the translation-rotation
decoupling is examined by the ratio of center of mass density relaxation
time, calculated from $F_s^C(k,t)$, to the rotational relaxation time as $\tau_{2R_g}/\tau_2$ 
scaled with its value at $T=$ 1.0 [see Fig. \ref{f:sedvp1}(d)]. Similar to the ratio $D\tau_2$,
the ratio $\tau_{2R_g}/\tau_2$ and  $D\tau_{2R_g}$ also oscillate around 1.0 (all these values
are scaled with their corresponding value at $T=$ 1.0), which means that translational molecular 
relaxation time and rotational relaxation time are coupled in the higher density system. However,
the ratio $\tau_{2R_g}/\tau_2$ starts increasing from $T=$ 0.9 and reaches around a value 1.25 
for $\rho=$ 0.85 and $\rho=$ 0.7, at the low temperature $T=$ 0.4. This shows a weak
decoupling between $\tau_{2R_g}$ and $\tau_2$ in both lower density systems. Such decoupling was
also found in an experimental study by Edmond \textit{et al.} \cite{j:kazem} in a colloidal glass,
and a simulation study of glassy dumbbells \cite{s:chong_kob}. We expect a strong violation of SE 
and SED relations in our system, if the observed trend is continued below $T_c$.

As many studies show that the violation of SED relation is attributed to the enhanced 
hopping process in the rotational motion \cite{j:micheleleporini2}. Therefore, we 
look into the typical trajectory of an orientation vector $\boldsymbol {\hat e}(t)$ over
the unit sphere of a few randomly selected polymer chains. In Fig. \ref{f:or}, we compare
the rotation of $\boldsymbol {\hat e}(t)$ of polymer chains at one higher temperature and
two lower temperatures. At $T=$ 1.0 of the density $\rho=$ 0.85, the trajectory 
of $\boldsymbol {\hat e}(t)$ [see Fig. \ref{f:or}(a)] shows a random walk, which uniformly
spans over the sphere. However, the higher and lower density systems at the low
temperatures [see Figs. \ref{f:or}(b--d)], show an intermittent motion of the
rotational vector, which is a bit pronounced in the lower density systems.
Figure \ref{f:or}(d) shows a trajectory of a randomly selected chain of the higher 
density system, which shows weak confinement at $T=$ 0.36. Earlier studies by
Jose \textit{et al.} show the pronounced violation of SED relation in nematogens during
isotropic to nematic transition, which is due to the strong confinement of nematic ordering
in the orientation of the molecules \cite{bb:ppj1,bb:ppj2}. Such confinements at specified 
orientations are absent in this system. As polymer molecules are short chains, which are
nearly spherically symmetric, and require more confinement to show SED violation.


\section{\label{se:sac} Concluding remarks} 

The violations of Stokes-Einstein and Stokes-Einstein-Debye relations are 
defining characteristics of the glass transition, which are explained in terms
of dynamic heterogeneities arise due to jumplike motions of mobile particles, and dynamic 
caging of immobile particles in simulations of atomistic model glass-formers \cite{b:berthierdh}.
Another important class of glass-forming liquids is polymers that are difficult to crystallize.
Direct observation of  violation of SE and SED relations in  simulations of model polymers 
are rare because of connectivity through the bonds between monomers, and the microscopic processes
such as jumplike motions are difficult to detect \cite{j:helfrich1}. However, the indirect
observation of microscopic origins of violations of SE relation is examined in the earlier 
studies \cite{j:leporini,j:leporini6}. Due to extended shape of the polymers, we widen our study
of the supercooled polymers near the glass transition to the lower density systems, where available 
volume for polymer chains is abundant, especially near the surface of dilute gas and supercooled
liquid domains coexistence. Extensive molecular dynamics simulations of a linear Lennard-Jones 
polymer chains are performed at monomer number densities $\rho$ = 0.7, 0.85, and 1.0 from
$T=$ 2.0--0.36. For the first two densities, the system forms domains of dilute gas and supercooled
liquid, whereas the higher density system is homogeneous near the glass transition temperature $T_c$.
Monomer density relaxation properties from the $F_s(k,t)$ at different time origins in the 
supercooled phase coexistence compared with that at $\rho=$ 1 at the same temperature show that 
the density relaxation is independent of the time origin in the systems where dilute gas and 
supercooled liquid coexist near the glass transition. The collective relaxation properties
obtained from $G_d(r,t)$ shows that the gas-supercooled-liquid domains are stable within our 
simulation time. In these systems that differ in their density, we look for direct evidence of SE
violation in the density relaxation of monomers and center of mass of polymer chains, and the SED
violation of the polymer chains, near their MCT glass transition temperatures.

We show that monomer SE relation is violated for all three systems in the supercooled regime,
which is pronounced in both lower density systems. The pronounced violation of the SE relation 
in both lower density systems is caused by the structural inhomogeneities and resulting dynamical
heterogeneity due to the enhanced disparity in the monomer mobility in comparison
to the higher density system. In the temperature range $T=$ 2.0--0.6, the molecular 
SE relation is obeyed in the higher and both lower density systems. However, in the
supercooled regime, the higher density system obeys the molecular SE relation,
whereas both lower density systems weakly violate it. At the lowest temperature of all 
three densities, we identify a  hump at $r=\sigma$ and a peak at small $r$
in $G_s(r,t)$ that together show the monomer cages and jumplike motions in the monomer
movement, which was also shown in the earlier studies of glass-forming binary mixtures
\cite{j:starr_fst,j:pan_SE}. Further, our study shows that disparity in the mobility of 
the monomers caused by the structural inhomogeneities is more in both lower density systems
compared to the higher density system. Thus, we show that violations in the monomer and molecular 
SE relations are attributed to the presence of mobile and immobile particles, the jumplike
motions, and caging in this linear polymer system at temperatures near $T_c$. We also show that
the unit vector associated with the polymer chains undergoes confinement. Thus, there is a weak
violation in the SED relation for both lower density systems, supported by the intermittent 
motion found in the typical trajectory of the end-to-end unit vector. Our study also shows that 
the glass transition in the presence of static  structural inhomogeneities is very much similar 
to the continuous phases. Many aspects of the formation of the glassy domains in the model glassy
binary mixtures are studied in the simulations \cite{j:testard1,anna:ppj1} 
and experiments \cite{j:tanakavps,j:frederic,j:godfrin}, where glass transition and phase
separation coexist, though the inter-molecular potentials and molecular geometry are
different from our study. Therefore, the simulations with more model potentials and 
varying range of attractions are required for obtaining the quantitative information about
the microscopic glassy domains formed in the phase separating systems.

\begin{acknowledgments}
We thank the HPC facility at IIT Mandi for computational support. PPJ acknowledges financial
support from SERB project no. EMR/2016/005600/IPC. The data that support the findings of this 
study are available from the corresponding author upon reasonable request.
\end{acknowledgments}

\appendix 

\renewcommand{\thefigure}{A.\arabic{figure}}
\setcounter{figure}{0}

\section{\label{sa:rg} Radius of gyration}
\begin{figure}
	\includegraphics[width=7.0cm, height=5.5cm]{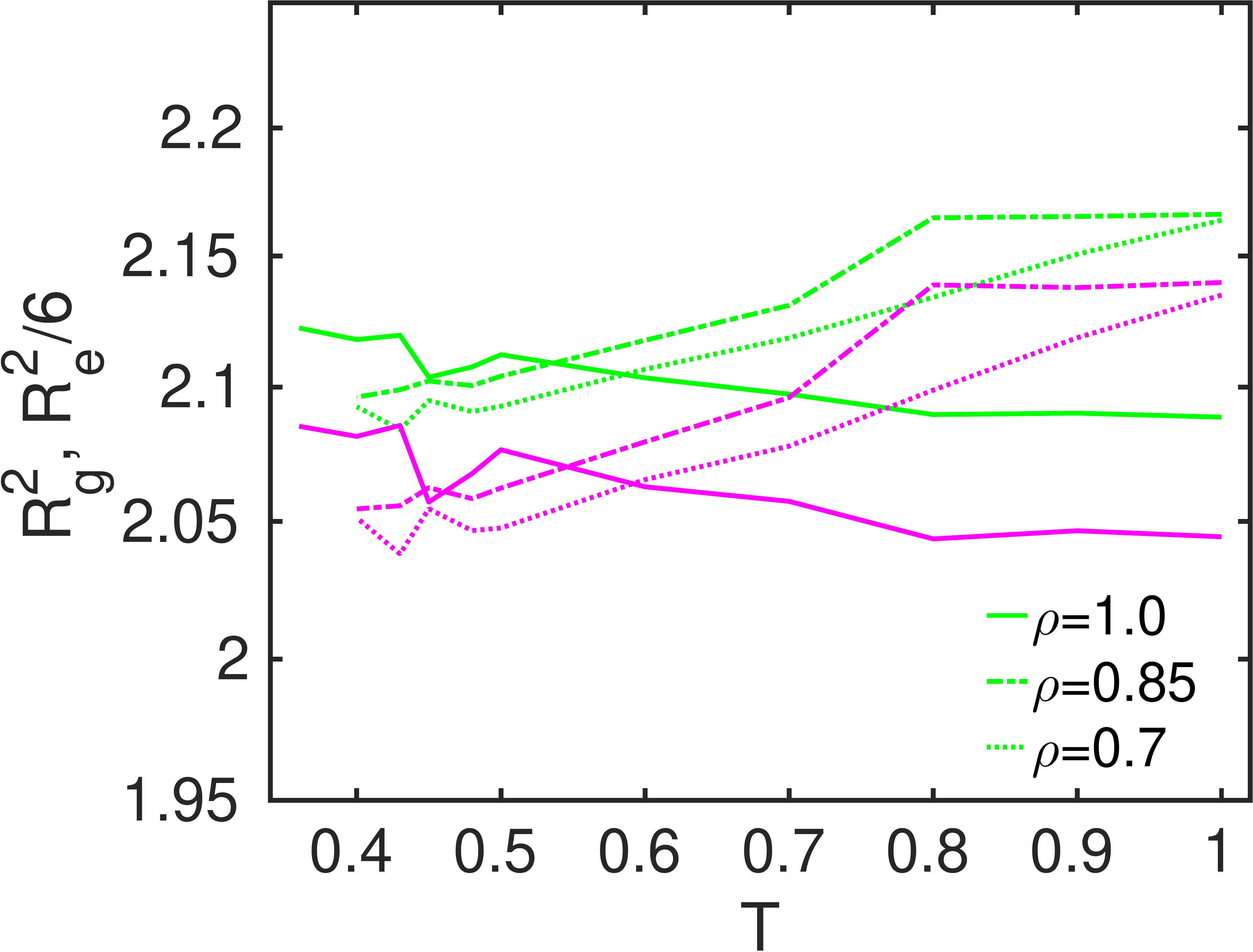}
	\caption{\label{f:rg} Average squared radius of gyration 
	$R^2_g$ (green) and $R^2_e/6$ (magenta) of $\rho=$ 0.7, 0.85,
	and 1.0 systems are plotted against temperature $T$.}
\end{figure}

The size of polymer chains can be measured from the 
calculation of radius of gyration. The equilibrium average mean square 
radius of gyration is
\begin{equation}
	{R^{2}_{g}} = {\frac{1}{n}}\left\langle\sum\limits_{i=1}^{n}{{({\mathbf{R}}_{i}-
	{\mathbf{R}}_{cm})}^{2}}\right\rangle,
\end{equation} 
where ${\mathbf{R}}_{cm} = {\frac{1}{n}} \sum\limits_{i=1}^{n}{{\mathbf{R}}_{i}}$
is a position of the center of mass of a chain and ${\mathbf{R}}_{i}$ is the position 
vector of  $i$th monomer and $n$ is the total number of monomers in a 
chain \cite{dae}. We show that (flexible) polymer chains are nearly Gaussian
as $R_g^2\simeq R_e^2/6$ (see Fig. \ref{f:rg}), partially because the 
chains are short therefore they are confined to an average small radius. 
Here, $R_g^2$ and $R_e^2$ are respectively radius of gyration and end-to-end 
distance squares. Calculations presented here show that the radius of gyration
changes with temperature by $\sim$ 2\% in the higher density system and 
$\sim$ 3\% in both lower density systems.

\section{\label{sa:packing} Local packing of monomers}
\begin{figure}
	\includegraphics[width=8.5cm, height=4.5cm]{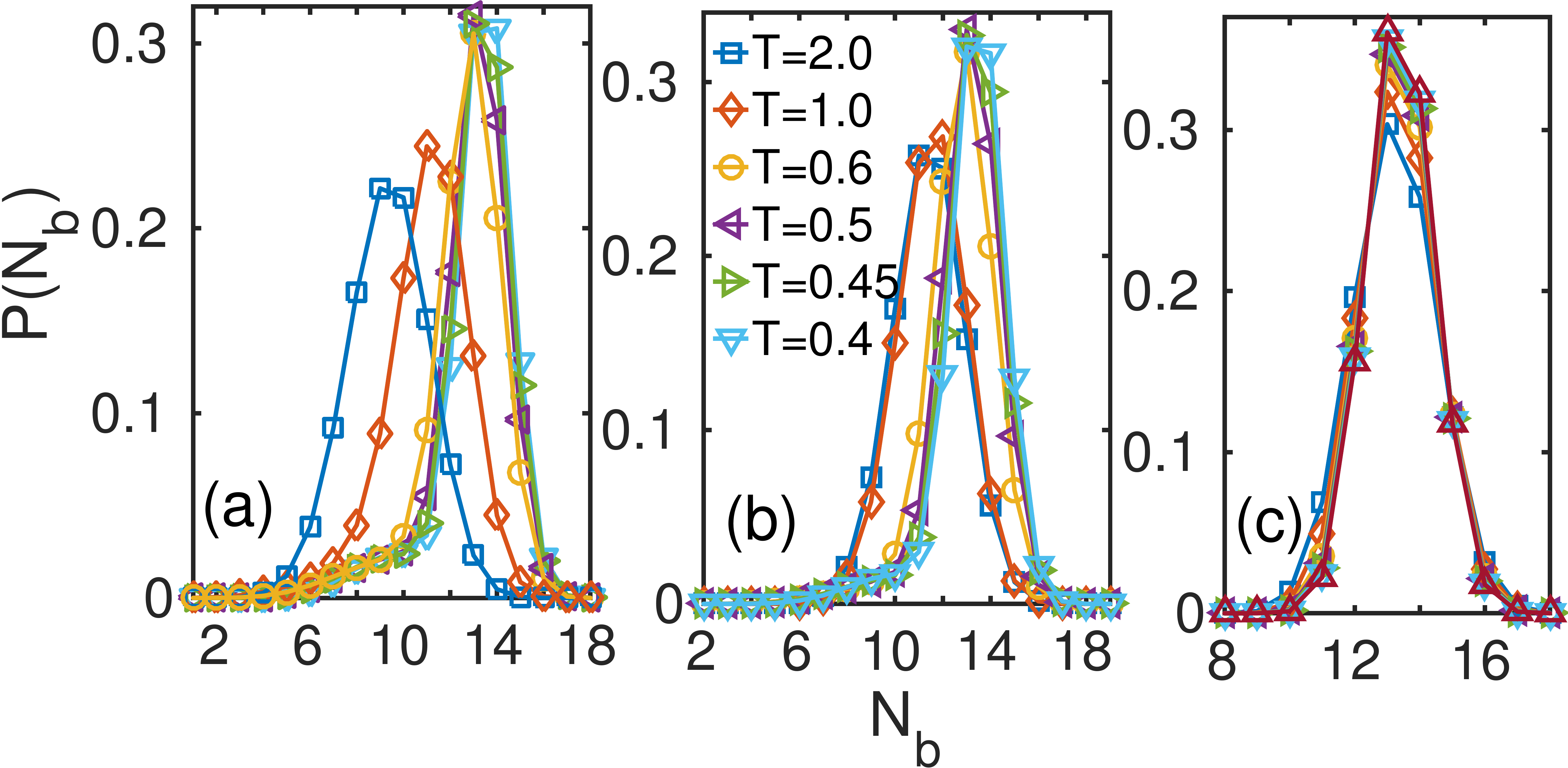}
	\caption{\label{f:pnb} Average probability distribution $P(N_b)$ is plotted against
	$N_b$ at $T=$ 2.0--0.36: (a) $\rho=$ 0.7, (b) $\rho=$ 0.85, and (c) $\rho=$ 1.0.
	In (a) and (b), $P(N_b)$ shows nonzero values for the smaller $N_b$, which corresponds to
	the gas-liquid phase coexistence at high temperatures. The liquid phase becomes amorphous
	solid-like coexisting with the gaseous particles at low temperatures.}
\end{figure}

Figure \ref{f:msdmap} and our previous study of this system \cite{jalim:ppj2} show that
lower density systems phase separate at low temperatures. To examine a variation
of the local density with temperature in the phase separating systems, we calculate local
packing of the monomers in the linear polymer system at three different densities.
The local packing is obtained from the calculation of number of nearest neighbors $N_b$ 
of each monomer in the first coordination shell (FCS), which is at 0 $<r\le r_{fcs}$; 
$r_{fcs}=$ 1.5 is a position of the first minima of the radial distribution function of 
the system \cite{jalim:ppj1,jalim:ppj2} . Using this FCS radius, its volume can be 
calculated as $V_{fcs}=(4/3)\pi r_{fcs}^3=$ 14.137, which subsequently gives the 
local density $\rho_l=(N_b+1)/V_{fcs}$. Figure \ref{f:pnb} shows that range of $N_b$ 
is 3--18 at $T=$ 0.4 for $\rho=$ 0.7 and 0.85, whereas $N_b=$ 9--18 at $T=$ 0.36 of the
higher density system. Thus, the local density range is calculated as
$\rho_l=$ 0.283--1.344 at $T=$ 0.4 of the lower densities $\rho=$ 0.7 and 0.85,
and $\rho_l=$ 0.707--1.344 at $T=$ 0.36 of the higher density system. This variation 
in the local density range shows a coexistence of dilute gas and dense amorphous
domains in both lower density systems at temperatures near $T_c$.

Further, the nearest neighbor distribution averaged over the steady-state
configurations [given in Figs. \ref{f:pnb}(a--c)] shows that the higher density system
does not show macroscopic cavities, whereas both lower density systems show cavities as
temperature reduces, resulting in the structural inhomogeneities \cite{jalim:ppj2}.
In both lower densities [see Figs. \ref{f:pnb}(a) and \ref{f:pnb}(b)], major peak of 
$P(N_b)$ shifts to $N_b=$ 13 and 14 at low temperatures, whereas in the higher density 
system [see Fig. \ref{f:pnb}(c)], the peak height grows at $N_b=$ 13 and 14 from high to
low temperatures. Thus, the packing of the monomers that are inside the dense domains are 
similar for all three densities because of the monomers within the dense domains in both 
lower density systems have the same range of $N_b$ to that the higher density system.
At low temperatures, a hump in $P(N_b)$ at $N_b\simeq$ 8--10 is appearing
in both lower density systems, which is absent in the higher density system. This hump 
in $P(N_b)$ confirms the structural inhomogeneity in both lower density systems, however,
an approximately equal peak height at $N_b=$ 13 and 14 indicates the similarity in the
local packing of the dense glassy domains in all three systems at low temperatures.

\begin{figure}
	\includegraphics[width=6.5cm, height=5.0cm]{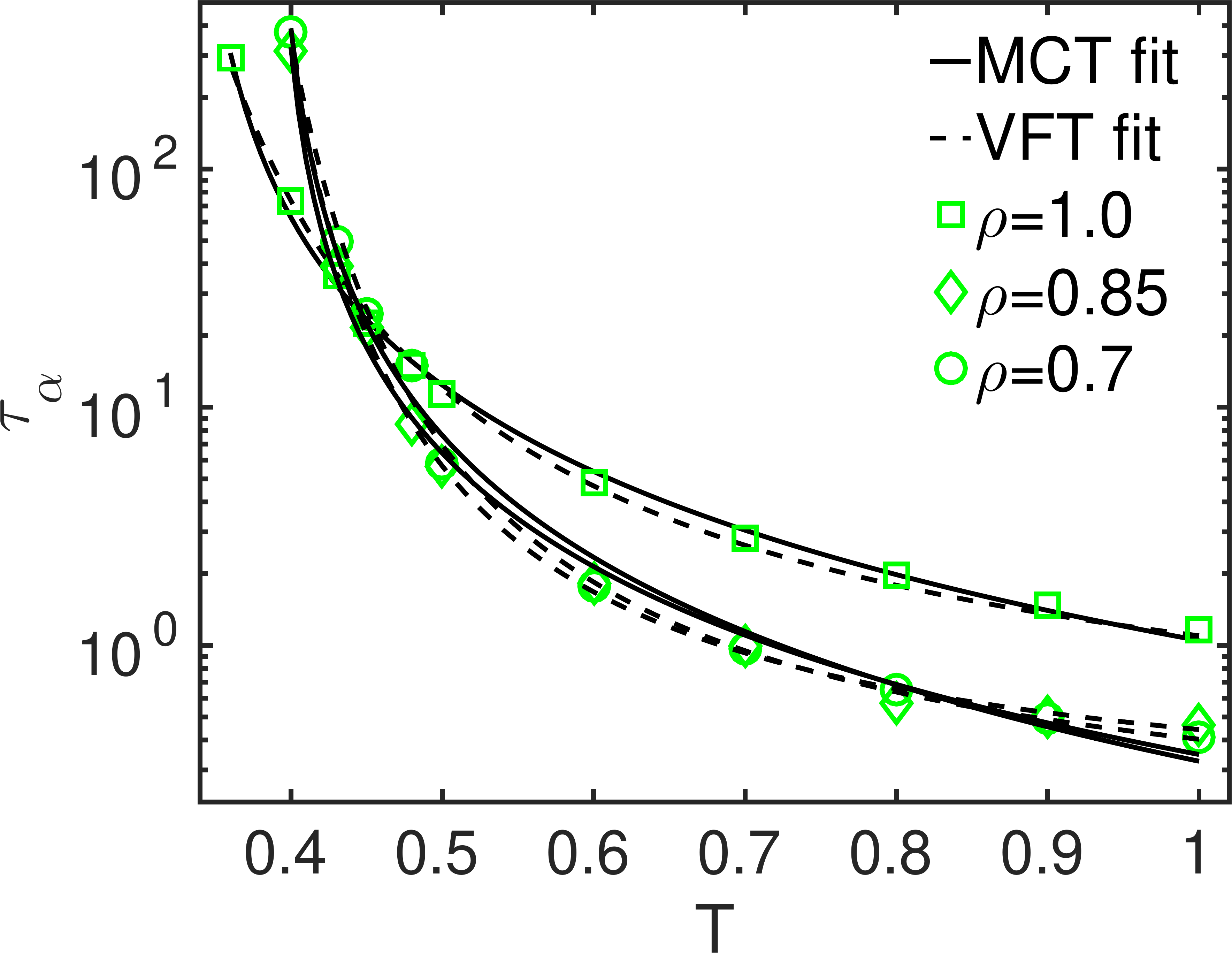}
	\caption{\label{f:talffit} Fitting of $\alpha-$relaxation time
	with the MCT and VFT relations. Note that the VFT relation is a good fit to the data.}
\end{figure}
\begin{figure}
	\includegraphics[width=7.5cm, height=6.0cm]{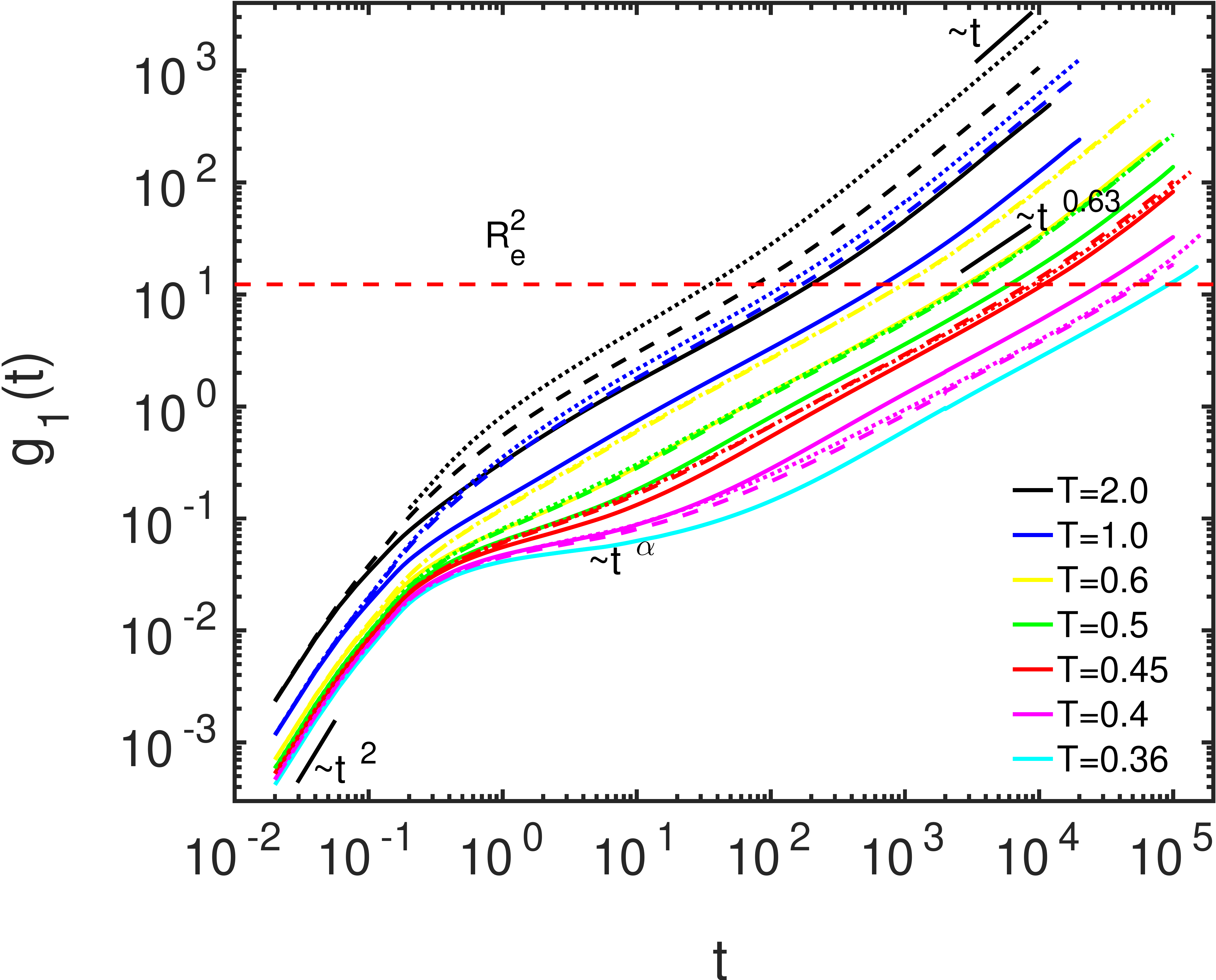}
	\caption{\label{f:msd} Mean squared displacement of monomers is plotted against
	time $t$. Solid, dashed, and dotted lines are corresponding to the densities
	$\rho=$ 1.0, 0.85, and 0.7 respectively.}
\end{figure}
\begin{figure}
	\includegraphics[width=7.0cm, height=8.5cm]{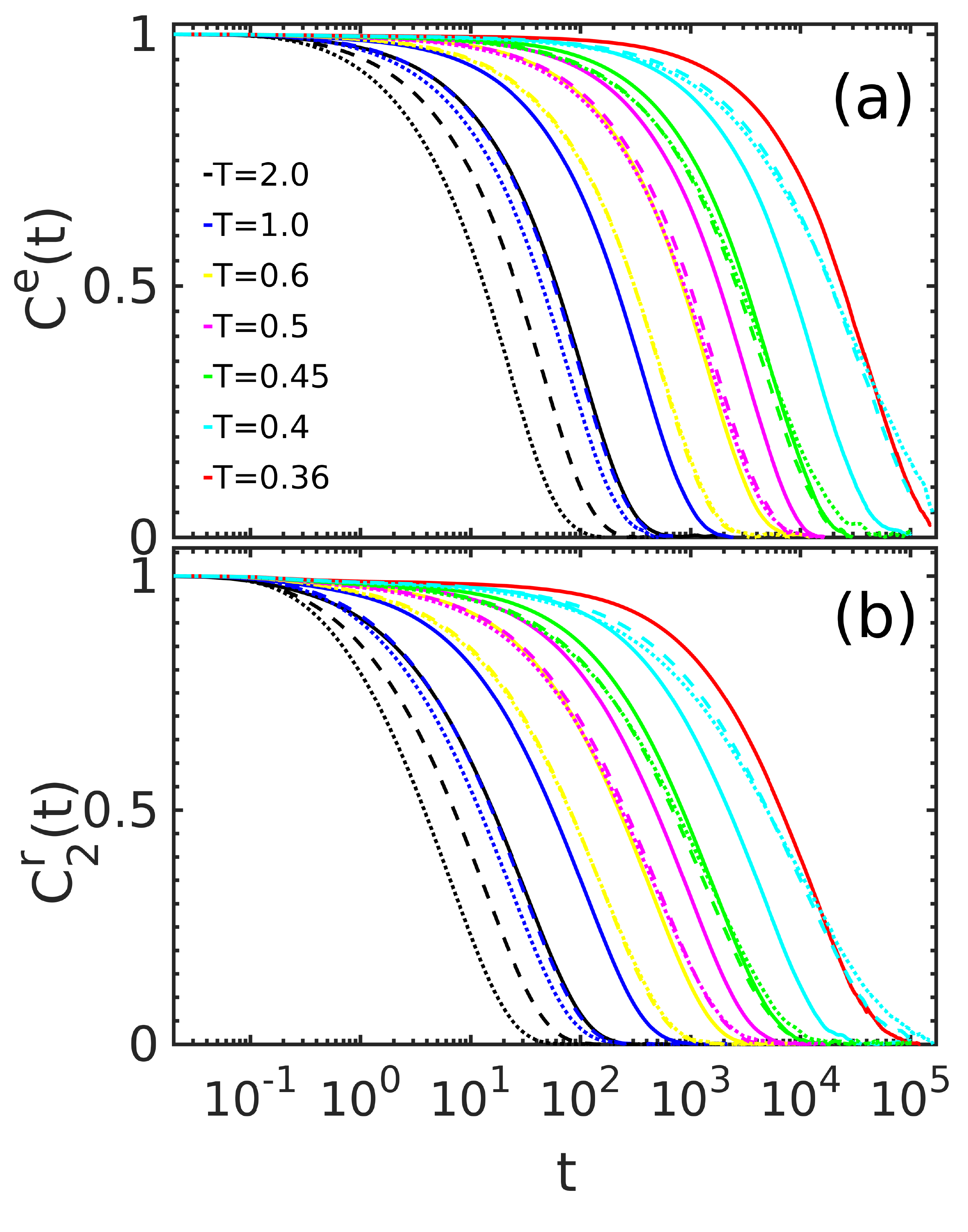}
	\caption{\label{f:eeor} End-to-end vector time correlation function (a) and Relaxation
	of second order time correlation function of the rotational vector (b) of polymer 
	chains. Solid, dashed, and dotted lines are corresponding
	to $\rho=$ 1.0, 0.85, and 0.7.}
\end{figure}
\begin{figure}
	\includegraphics[width=8.0cm, height=5.0cm]{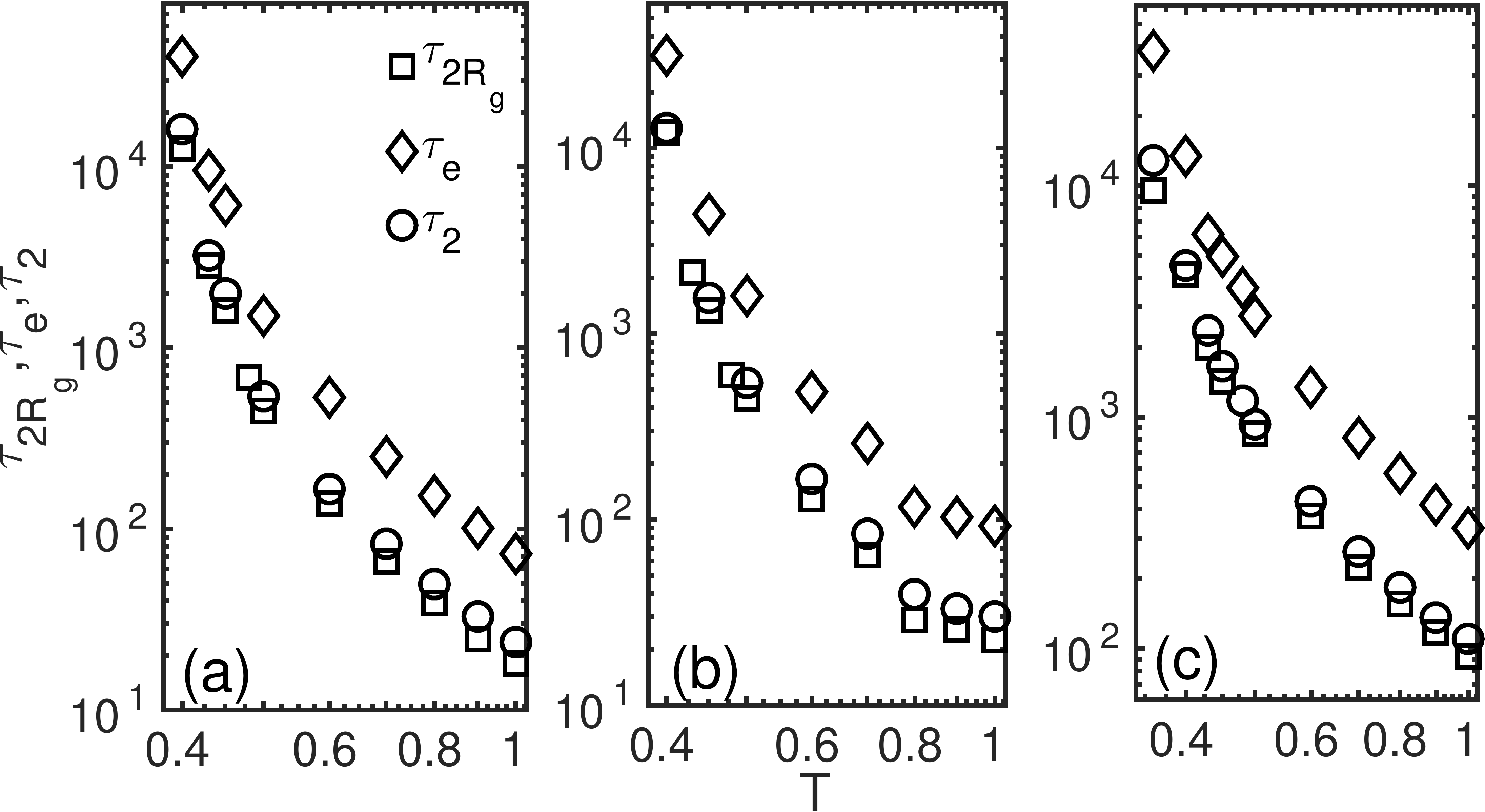}
	\caption{\label{f:orlxtym} Molecular relaxation times $\tau_{2R_g}$, $\tau_e$, and 
	$\tau_2$ are plotted against temperature $T$ at densities (a) $\rho=$ 0.7, (b) $\rho=$ 0.85,
	and (c) $\rho=$ 1.0.}
\end{figure}
\begin{figure}
	\includegraphics[width=8.0cm, height=6.5cm]{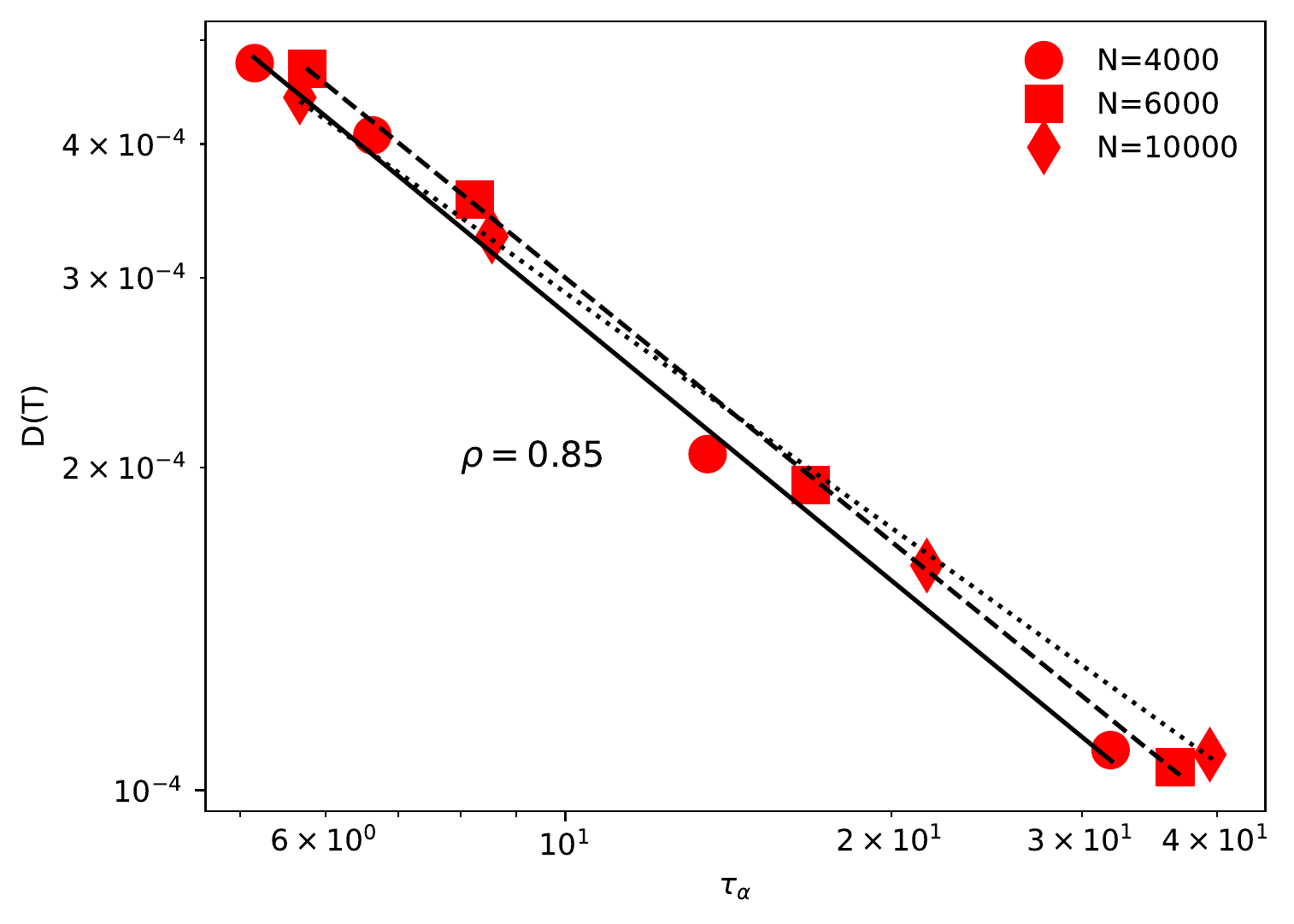}
	\caption{\label{f:finsizeSE} Diffusion coefficient $D(T)$ is plotted against monomer 
	$\alpha-$relaxation time at different system sizes, in the range of
	temperatures ($T=$ 0.5--0.43) where the SE relation is violated. Data is fitted with the 
	fractional SE relation, i.e., $D\propto\tau^{-\xi}$. Solid, dashed, and dotted lines 
	are fit to the data corresponding to $N=$ 4000, 6000, and 10000. The corresponding
	exponents are 0.83, 0.82, and 0.73.}
\end{figure}
\begin{figure}
	\includegraphics[width=8.7cm, height=12.7cm]{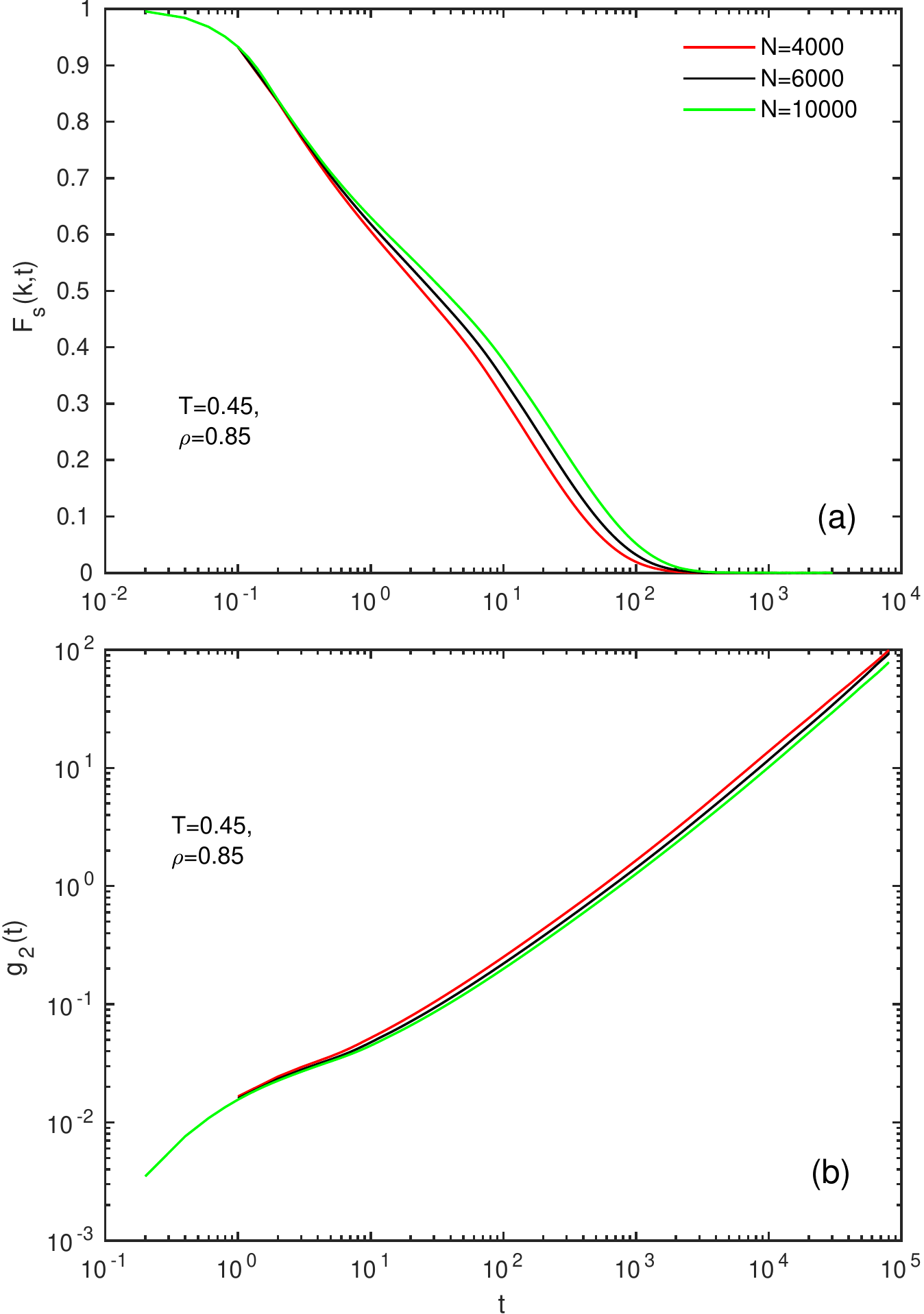}
	\caption{\label{f:finsizeDyn} Slow dynamics of different system sizes at one 
	representative temperature $T=$ 0.45. (a) $F_s(k,t)$ and (b) Center of mass MSD, $g_2(t)$.
	This shows that $F_s(k,t)$ and center of mass MSD slow down as the system size increases.}
\end{figure}
\begin{figure}
	\includegraphics[width=8.0cm, height=6.5cm]{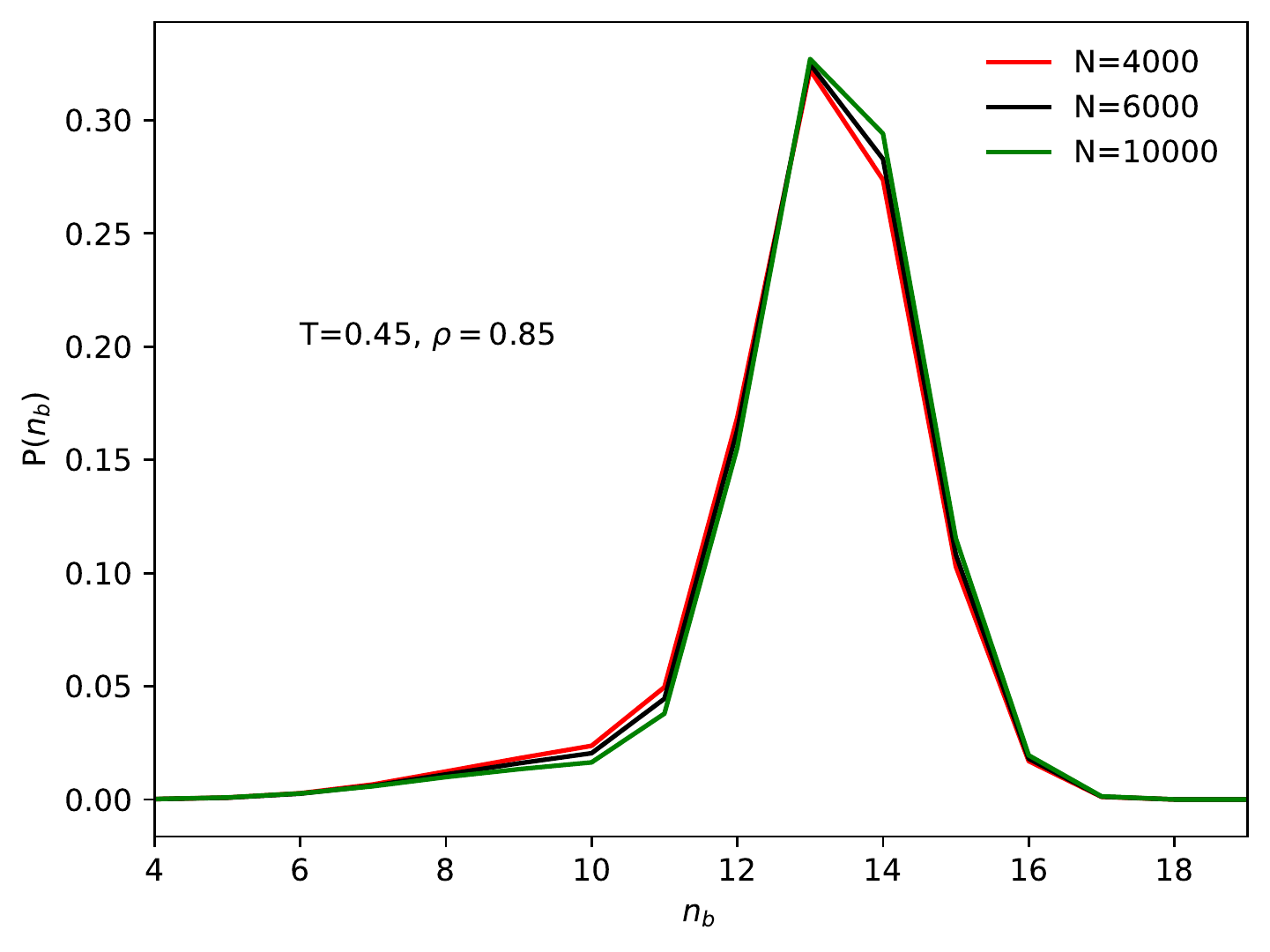}
	\caption{\label{f:finsizePnb} Nearest neighbor distribution, $P(n_b)$ at different system 
	sizes of one representative temperature $T=$ 0.45. $P(n_b)$ at low
	$n_b$ decreases whereas it increases at $n_b=$ 14, with the system sizes.}
\end{figure}

\section{Relaxation time and the fitting\label{sa:taufit}}
The monomer $\alpha-$relaxation times are plotted against temperature $T$ in
Fig. \ref{f:talffit} at three different densities. The $\tau_\alpha$ \textit{vs} $T$ 
curves are fitted with the MCT and VFT relations and described in the main text. 
The $\alpha-$relaxation time is longer for the higher density system compared to
both  lower density systems up to $T=$ 0.45. At $T=$ 0.4, the $\alpha-$relaxation 
time of both lower densities show a crossover to the higher density system, 
which is also evident from the slower decay of the $F_s(k,t)$. 

\section{Monomer mean-squared displacement \label{sa:trans}}
An average squared displacement of a particle from its initial position can be calculated 
from its mean-squared displacement (MSD)
\begin{equation}
	g_1(t)=\langle \left[\mathbf{r}(0) - \mathbf{r}(t) \right]^2  \rangle,
\end{equation}
which is shown in Fig. \ref{f:msd} at various temperatures of the three different
densities. Figure \ref{f:msd} shows that monomers show the ballistic motion ($\sim t^2$) at
a short time, crossing over to the intermediate caging 
regime ($\sim t^\alpha$, $0<\alpha<1$) owing to various interactions, bonding as
well as non-bonding type. There exists a sub-diffusive regime in the MSD of monomers after
the caging regime, which is due to the chain hindrance and varies with power-law 
$g_1(t)\sim t^{0.63}$ \cite{j:barratrev}. Finally, the diffusive regime in the monomer 
MSD starts from time $t\sim 10^3$ at the temperature $T=$ 2.0, whereas at lowest temperatures
($T=$ 0.4 and $T=$ 0.36) the diffusive regime starts at the time scale 
of $t\sim 10^6$ \cite{j:barratrev}. As the temperature of the system gets lowered the caging
of the monomer MSDs becomes more pronounced, which is supported by the $F_s(k,t)$ shown in 
the main text. The monomer MSD of the higher density system is slower than 
both lower density systems above $T=$ 0.4, shows a crossover at $T=$ 0.4 where the monomer
MSD of both lower density systems becomes slower. 

\section{Orientation\label{sa:orr}}
Further understanding of the molecular relaxation is obtained from the rotational
motion of the polymer chains. An autocorrelation function of the End-to-end vector
is defined as \cite{dae}
\begin{equation}
	C^{e}(t) = \frac {\left\langle\mathbf e(0).\mathbf e(t)\right\rangle}
	{\left\langle\mathbf e(0).\mathbf e(0)\right\rangle},
\end{equation}
which gives a detail of the molecular shape relaxation in the polymers. Figure \ref{f:eeor}(a) 
shows the $C^{e}(t)$ of polymer chains at different temperatures of three densities. The 
relaxation of $C^e(t)$ is slower with an increase in the density at
temperatures $T=$ 2.0 and 1.0. Below these temperatures, the relaxation of $C^e(t)$ 
becomes identical for both lower densities. However, $C^e(t)$ of the higher
density system decays slower than both lower density systems up to $T=$ 0.45. A 
crossover in the relaxation time of both lower density systems appears at $T=$ 0.4, and 
$C^e(t)$ relaxes slowly. We have computed the time constant for end-to-end vector relaxation
at all temperatures, which is defined as $ \tau_e =\int_0^\infty C^{e}(t)dt$. A variation 
in $\tau_e$ with $T$ is plotted in Fig. \ref{f:orlxtym} at three different densities, which
is qualitatively similar to the $C^{e}(t)$. The relaxation of $C^{e}(t)$ has contributions
from the rotation as well as shape fluctuations of the polymer chains, thus we compute 
rotational correlation functions. 

The rotational relaxation time is quantified from the rotational correlation function
$C_l^r(t)$ for a non-spherical molecule \cite{bap}. Hydrodynamic SED model of rotational
relaxation predicts an exponential relaxation of $C_l^r(t)$, i.e.,
$C_l^r(t) = e^{-t/\tau_l}$ for liquids at high temperatures, where $\tau_l$ is the time
constant of $l^{th}$ order orienational correlation function. The rotational correlation 
function, corresponding to $l=$ 2, i.e., $C_2^r(t)$, is plotted in Fig. \ref{f:eeor}(b).
The effect of density on the nature of the relaxation of $C_2^r(t)$ is similar to the $C^e(t)$,
as described above. At high temperatures, $C_2^r(t)$ shows exponential relaxation as predicted
by the SED model. However, at temperatures ($T=$ 0.45--0.4), $C_2^r(t)$ not only slows down but also
shows an emergence of a shoulder, immediately after the fast initial decay. The appearance of the
shoulder hints the formation of cages in the rotational motion of the polymer molecules due to the 
orientational confinement at small angles. These small-angle confinements are correlated to the
confinement of rotational motion, shown in the trajectory of the unit
vectors (see Fig. \ref{f:or}).

A comparison of variation in translational molecular relaxation time $\tau_{2R_g}$, end-to-end 
vector relaxation time $\tau_e$, and rotational relaxation times $\tau_2$ with 
temperature and density, is shown in Fig. \ref{f:orlxtym}. All these relaxation 
times grow as temperature reduces for all three density systems, though they are higher for the
higher density system up to $T=$ 0.45. A crossover in the relaxation times is also observed
around $T=$ 0.4 for both lower density systems. Interestingly, these relaxation time at 
$T=$ 0.36 of the higher density system, reach near to  the value at $T=$ 0.4 of both
lower density systems.

\section{Finite-size effects\label{sa:finsize}}

To examine finite-size effects due to cavities, we have studied two smaller size systems at the density
$\rho=$ 0.85, consisting of $N=$ 4000 and 6000 number of particles (beads), and compared results
with that of the system of $N=$ 10000 particles. A plot of $D$ \textit{vs} $\tau_\alpha$ at different
system sizes, given in Fig. \ref{f:finsizeSE}, shows that in all three systems the relaxation time 
increases with system size, and the diffusion coefficient decreases by a smaller factor, which 
results in a variation in the violation of the SE relation with system size. The exponent $\xi$ 
decreases with increasing system size, indicating that the larger system shows the larger violations.
To analyze how relaxation time and diffusion vary with system size, the $F_s(k,t)$ of monomers
and center of mass MSD of polymer molecules, at a representative temperature $T=$ 0.45, are compared
at different system sizes in Figs. \ref{f:finsizeDyn}(a-b), showing a slow down of strucutral relaxation 
with an increase of the system sizes. This observed variation can be associated with a change in the 
relative distribution of surface and bulk particles in the system; this is identified from the 
nearest-neighbor distribution, shown in Fig. \ref{f:finsizePnb}, where it shifts to the higher $n_b$ 
as the system size increases. This leads to the larger disparity in the mobility of particles, thus
enhances the extent of the SE violations. It is expected that the finite-size effects may saturate at very 
large system sizes which may be worthwhile to address in future studies using simulations or experiments. 
As there is an enhancement of SE violations with system size, experiments that involve a larger number of
particles are expected to show a more pronounced SE violations.

\bibliography{poly}

\begin{thebibliography}{85}
\bibitem [{sen()}]{senotes}%
  \BibitemOpen
  \href@noop {} {}\bibinfo {note} {In liquids at high temperatures,
  intermediate incoherent scattering function can be given as
  $F_s(k,t)=\exp(-Dk^2t) \equiv \exp(-t/\tau)$ \cite{ham}, here, $k$ is a wave
  vector near the first peak of static structure factor $S(k)$. This suggests
  that $Dk^2 \tau = $constant, or $D\tau = $constant, implies that $D \propto
  1/\tau$, which according to $D\eta/T=$ constant, entails that $\tau \propto
  \eta/T$.}\BibitemShut {Stop}%
\end{thebibliography}

\end{document}